\documentclass[journal]{IEEEtran}
 \usepackage{amsmath,amssymb}
 \usepackage{subfigure}
 \usepackage{graphicx,graphics,color,psfrag}
 \usepackage{cite,balance}
 \usepackage{algorithm}
 \usepackage{accents}
 \usepackage{amsthm}
 \usepackage{bm}
 \usepackage{url}
 \usepackage{algorithmic}
 \usepackage[english]{babel}
 \usepackage{multirow}
 \usepackage{enumerate}
 \usepackage{cases}
 \usepackage{stfloats}
 \usepackage{dsfont}
 \usepackage{color,soul}
 \usepackage{amsfonts}
 \usepackage{cite,graphicx,amsmath,amssymb}
 \usepackage{subfigure}
 \usepackage{fancyhdr}
 \usepackage{hhline}
 \usepackage{graphicx,graphics}
 \usepackage{array,color}
 \usepackage{amsmath}
 \usepackage{amsthm}

\newtheorem{definition}{Definition}

\newtheorem{lemma}{Lemma}

\newtheorem{proposition}{Proposition}
\newtheorem{remark}{Remark}

\def\qed{$\Box$}

\def\proof{\noindent{\emph{Proof:} }}

\def
\endproof{\hspace*{\fill}~\qed
\par
\endtrivlist\unskip}
\def\endproof{\hspace*{\fill}~\qed\par\endtrivlist\vskip3pt}

\def\E{\mathsf{E}}

\def\l{\left}
\def\r{\right}
\def\({\left(}
\def\){\right)}

\def\bff{{\mathbf{f}}}

\def\bh{{\mathbf{h}}}

\def\bn{{\mathbf{n}}}

\def\bs{{\mathbf{s}}}

\def\bu{{\mathbf{u}}}

\def\bw{{\mathbf{w}}}

\def\b0{{\mathbf{0}}}

\def\bA{{\mathbf{A}}}
\def\bB{{\mathbf{B}}}

\def\bF{{\mathbf{F}}}
\def\bG{{\mathbf{G}}}
\def\bH{{\mathbf{H}}}
\def\bI{{\mathbf{I}}}

\def\bQ{{\mathbf{Q}}}

\def\bU{{\mathbf{U}}}
\def\bV{{\mathbf{V}}}
\def\bW{{\mathbf{W}}}
\def\bX{{\mathbf{X}}}
\def\bY{{\mathbf{Y}}}

\newcommand{\nn}{\nonumber}

\DeclareMathSizes{10}{9.6}{5}{3}

\def\papertitle{ \Huge MIMO Over-the-Air Computation \\ for High-Mobility Multi-Modal Sensing}

\IEEEoverridecommandlockouts 

\begin{document}

\title{ \fontsize{21}{21}\selectfont \papertitle}
\author{Guangxu Zhu and Kaibin Huang
\thanks{ G.~Zhu and K.~Huang are with the Dept. of EEE of The University of Hong Kong (HKU), Pok Fu Lam, Hong Kong.  Corresponding author: K. Huang (Email:   haungkb@eee.hku.hk). }
}
\maketitle

\begin{abstract}
 In future \emph{Internet-of-Things} (IoT) networks, sensors or even access points can be mounted on ground/aerial vehicles for smart-city surveillance or environment monitoring. For such  \emph{high-mobility sensing}, it is impractical to collect data from a large population of sensors using any traditional orthogonal multi-access scheme as it would  lead to excessive latency. To tackle the challenge, a  technique called \emph{over-the-air computation} (AirComp) was recently developed to enable a data-fusion to receive a desired function (e.g., averaging or geometric mean) of sensing data from concurrent sensor transmissions. This is made possible by exploiting the superposition property of a multi-access channel. 
Targeting a  multi-antenna sensor network,   this work aims at developing \emph{multiple-input-multiple output} (MIMO) AirComp for enabling \emph{high-mobility multi-modal  sensing}  where a multi-modal sensor monitors multiple environmental parameters such as temperature, pollution and humidity.  To be specific, we design MIMO-AirComp equalization and channel feedback techniques for spatially multiplexing multi-function computation, each corresponding to a particular sensing-data type.  Given  the objective of minimizing sum mean-squared error via spatial diversity, a close-to-optimal equalizer is derived in closed-form  using differential geometry.  The solution can be computed as the weighted centroid of points (subspaces) on a  Grassmann manifold, where each point represents the subspace spanned by the channel coefficient matrix of a sensor.  As a by-product, the problem of  MIMO-AirComp equalization  is proved to have the same form as the classic problem of multicast beamforming, establishing the  \emph{AirComp-multicasting duality}. Its significance lies in making the said Grassmannian-centroid solution method transferable to the latter problem which otherwise is solved using the more computation-intensive \emph{semidefinite relaxation} method in the literature.   Last, building on the AirComp equalization  solution, an efficient channel-feedback technique is designed for an access point to receive the equalizer  from simultaneous sensor transmissions of designed signals that are functions of local channel-state information. This overcomes the difficulty of provisioning orthogonal feedback channels for many sensors. 
 
 \end{abstract}


\IEEEpeerreviewmaketitle

\section{Introduction}

\begin{figure*}[tt]
  \centering
  \subfigure[Smart-city surveillance with UAV mounted  access point]{\label{Fig:uav_ap}\includegraphics[width=0.46\textwidth]{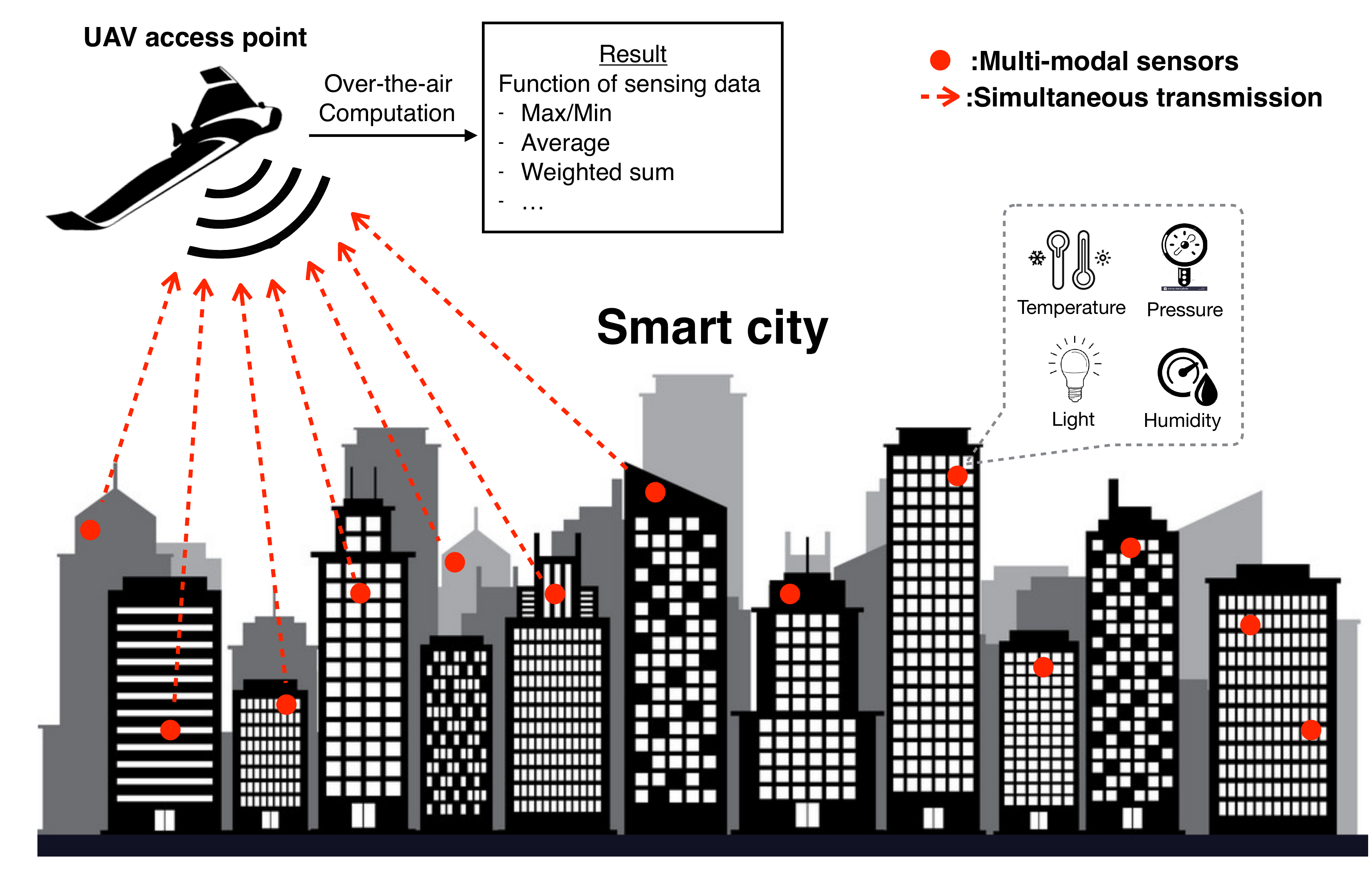}}
  \hspace{0.35in}  
  \subfigure[Wild-area  monitoring with UAV mounted  sensors]{\label{Fig:uav_sensor}\includegraphics[width=0.46\textwidth]{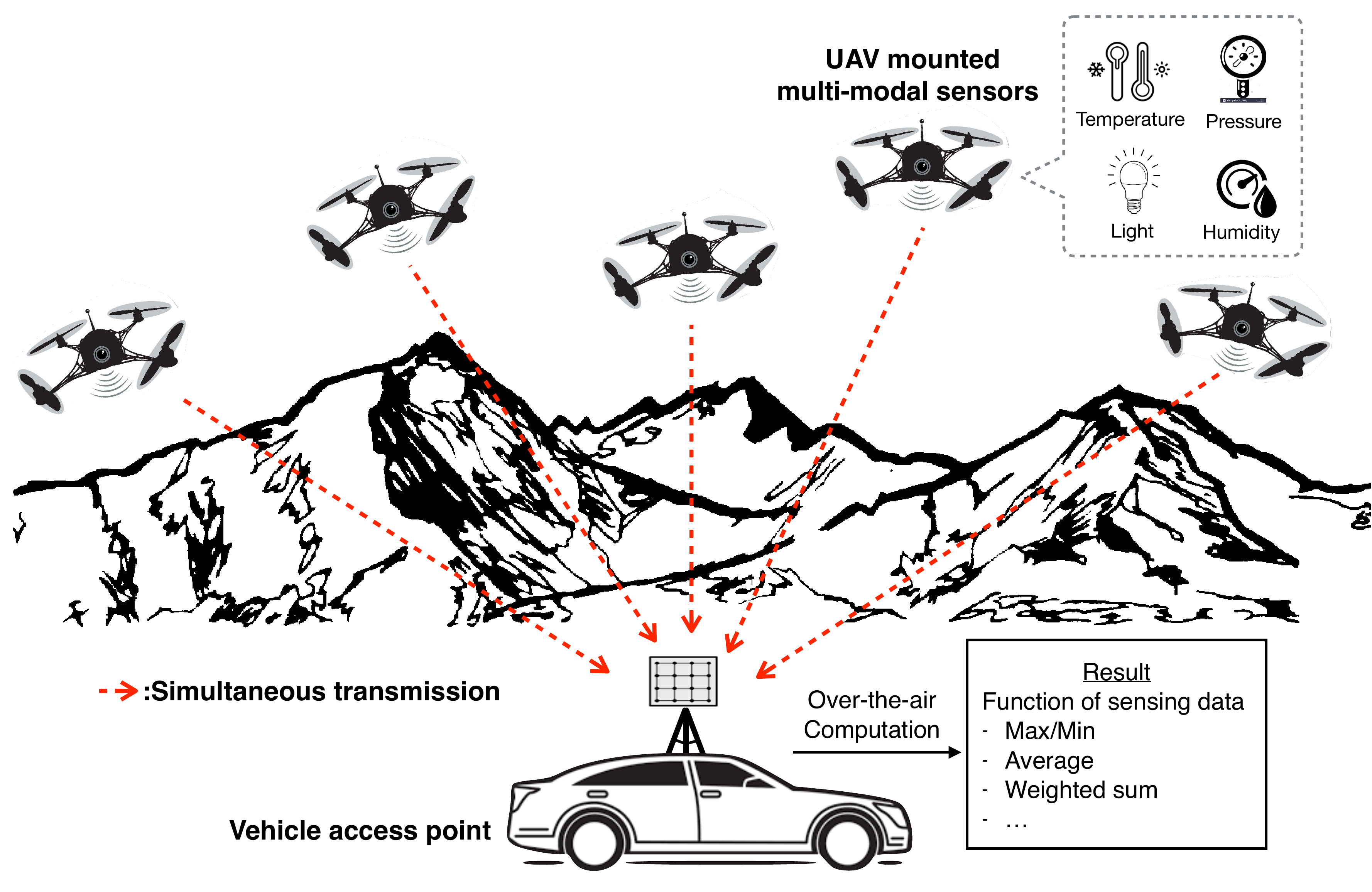}}
  \caption{Two scenarios for   high-mobility multi-modal sensing.}
  \label{model1:subfig}
  \vspace{-4mm}
\end{figure*}

Attaining  the vision of \emph{Internet-of-Things} (IoT) will require the ubiquitous deployment of an enormous number of sensors  (e.g., tens of billions) in our society \cite{AgiwalTutorial2016,zhu2018inference}. The brute-force  approach of ``transmit-then-compute" is obviously impractical for this  large-scale sensor network  as the massive radio access would result in  excessive network latency and low efficiency in spectrum utilization. 
The situation is exacerbated  at high mobility where ultra-fast data aggregation from many sensors is desired. This is the case when 
sensors and/or the \emph{access-point} (AP) are mounted on  ground vehicles or \emph{unmanned  aerial vehicles} (UAV)  for ubiquitous city surveillance  in  the smart-city application  [see Fig. \ref{Fig:uav_ap}], or for wild-area  monitoring to avoid natural disasters  [see Fig. \ref{Fig:uav_sensor}].
Motivated by the need of ultra-fast data aggregation, an intelligent solution, known as \emph{over-the-air computation} (AirComp), is proposed recently that  exploits the signal-superposition property of a \emph{multi-access channel}  (MAC) to compute a class of so called \emph{nomographic}  functions  of  distributed sensing data via concurrent sensor transmissions (see Fig. \ref{model1:subfig}), thereby integrating computation  and communication  \cite{KatabiAirComp2016, GastparTIT2007}.  Examples of such functions are show in Table \ref{summary:table1}. Unlike rate-centric wireless systems where simultaneous transmissions result in interference, the computation accuracy for a sensor network with AirComp may grow with the number of simultaneous sensors due to the sensing-noise averaging. In this paper, we aim to advance   the area of AirComp  by developing \emph{multiple-input-multiple output} (MIMO) AirComp  for next-generation multi-antenna multi-modal sensor networks. The technology supports \emph{high-mobility multi-modal} (HMM) sensing by enabling  multi-function computation via spatial multiplexing and accurate  reception of the results by exploiting spatial diversity. 

\subsection{Over-the-Air Computation for Sensor Networks}
The idea of AirComp can be tracked back to  the pioneer work studying functional  computation in  sensor networks \cite{GastparTIT2007}.\footnote{AirComp also  appeared the same time as a key  operation  in  the  scheme  named  physical-layer  network coding  in \cite{zhang2006hot}.}  In \cite{GastparTIT2007},   structured codes (e.g., lattice codes) are designed for reliably computing  at an AP a function of distributed sensing values transmitted over  a MAC. The significance of the work lies in its counter-intuitive finding that ``interference" can be harnessed to help computing.  Subsequently, it was proved that the simple analog transmission without coding, where transmitted signals are scaled versions of  sensing values,  can  achieve the minimum functional distortion achievable by any scheme   \cite{GastparTIT2008}. On the other hand, coding can be still useful for other settings such as sensing correlated Gaussian sources   \cite{SoundararajanTIT2012}. The satisfactory performance (with optimality in certain cases) of simple \emph{analog AirComp} have led to an active area focusing on  designing and implementing techniques for  receiving  a desired function of concurrent signals, namely a targeted coherent combination of  the signal waveforms \cite{GoldsmithTSP2008,C.H.WangTSP2011,GoldenbaumICC2015,GoldenbaumWCL2014,GoldenbaumTCOM2013,GoldenbaumTSP2013,KatabiAirshare2015,KatabiAirComp2016,SiggICIT2012,Goldenbaumsensors2014}. In particular, considering analog AirComp,  power control at sensors was optimized in  \cite{GoldsmithTSP2008,C.H.WangTSP2011},  the computation rate (defined as the number of functional values computed per time slot) analyzed in \cite{GoldenbaumICC2015},  and the effect of channel estimation error  characterized in \cite{GoldenbaumWCL2014}. 



The implementation of  AirComp  faces several practical issues. One is the synchronization of all active sensors required for coherent combining at the AP. To cope with  synchronization errors, a scheme was proposed in \cite{GoldenbaumTCOM2013, GoldenbaumTSP2013} where the sensing value is modulated at each sensor  as the power of transmitted signal  and furthermore  a random phase rotation is applied to the  signal. The design transforms functional computation at the receiver to power detection while synchronization error appears as  random noise. An alternative solution, called \emph{AirShare},  was developed in \cite{KatabiAirshare2015}  for synchronizing sensors by broadcasting a  reference-clock  signal and its effectiveness was demonstrated using a prototype. In addition,  by applying appropriate data pre/post-processing, later work extended and implemented  AirComp  to compute  a variety of functions besides the linear ones  (as summarized in Table \ref{summary:table1}) \cite{SiggICIT2012,Goldenbaumsensors2014}.

It is worth mentioning that the coding techniques  designed for AirComp in  computing-centric sensor networks \cite{GastparTIT2007} inspired researchers to adopt relevant principles and ideas in designing new schemes for rate-centric  communication networks \cite{GastparTIT2011,T.YangTWC2014,L.ShiTIT2016,GastparTIT2014,SakzadTWC2013}. For relay assisted networks, the application of AirComp at relay nodes for  decoding  and forwarding  linear functions  of the transmitted messages led to the invention of the well known  \emph{compute-and-forward} relaying schemes \cite{GastparTIT2011,T.YangTWC2014,L.ShiTIT2016}.  Building on lattice coding, a novel so called integer-forcing linear receiver was proposed for spatial multiplexing in a \emph{multiple-input-multiple-output} (MIMO) system that attempts to create an effective channel matrix with integer coefficients to facilitate lattice decoding. The key operation, integer forcing, is similar to AirComp and computes a desired set of linear functions with  integer coefficients \cite{GastparTIT2014,SakzadTWC2013}.  In parallel with the above  research, extensive progresses were made in the area of physical layer network coding  where the celebrated network coding schemes invented for wired networks were  extended to wireless networks with AirComp relays (see survey in \cite{GastparProceedings2011}). 

Sensing devices targeting emerging applications such as smart cities are sophisticated. Each typically contains multiple multi-modal sensors monitoring different environmental parameters (e.g., pressure, light, humidity, and temperature) \cite{gubbi2013internet}. In particular, a smartphone recruited for crowd-sensing  typically contains  seven  or more sensors  such as inertial, GPS, and light sensors \cite{ganti2011mobile}. In view of prior work, the existing solutions  focus only  single-function AirComp assuming single-antenna sensors having uni-modal sensing capabilities. However, next-generation wireless networks equipped with large-scale arrays will make it possible to simultaneously compute multiple functions of multi-modal sensing data over-the-air. This inspires the current work on developing the technology of AirComp for a MIMO MAC, which can simultaneously spatial multiplex multi-function computation and suppress computation errors by exploiting spatial-diversity gain. Thereby, the data-fusion latency in sensor networks can be substantially reduced, meeting the ultra-low latency requirement in next-generation networks especially when high-mobility support is needed \cite{AgiwalTutorial2016}.

\subsection{MIMO Beamforming: Multi-Access versus AirComp}

Beamforming design for multiuser MIMO systems is a classic topic that has been extensively studied and there exists a rich relevant literature \cite{foschini1996layered,gesbert2007shifting, spencer2004introduction}.  In terms of network topology, the multi-antenna multi-modal sensor network we consider is equivalent to a MIMO multi-access communication network where a single AP supports simultaneous uplink transmissions of multiple users.  For the communication network, the designs of multiuser MIMO beamforming at the AP can be largely grouped into capacity-achieving  nonlinear designs based on  \emph{successive interference cancellation} (SIC) \cite{foschini1996layered} and low-complexity linear designs (\emph{minimum mean-squared error} (MMSE) or zero-forcing)  \cite{gesbert2007shifting,spencer2004introduction,zhu2017hybrid}. All of the designs share  the same objective  of decoupling multiuser signals by interference suppression and spatial multiplexing of data streams for each user. In contrast, AirComp receive beamforming in the sensor network has a different objective of minimizing the total distortion in the received values of multiple functions combining multi-modal data simultaneously sent by a set of sensors. Due to the difference in  objective between communication and sensing, the known designs for the former are inapplicable for the latter. On the other hand, existing AirComp literature considers only uni-function computation targeting  uni-modal sensing as discussed earlier. This makes receive beamforming for multi-function AirComp for multi-modal sensing an uncharted problem to be tackled in this work. 

It is worth mentioning that the discovery of \emph{uplink-downlink duality} is a breakthrough  in multiuser MIMO communication. The duality reveals similar structures in optimal beamformers for the MACs and broadcast channels that exist under various performance criteria ranging from capacity maximization \cite{jindal2004duality,Weiyu2006duality} to MMSE \cite{bjornson2014optimal}. This allows beamforming designs derived for the MACs to be applied to the broadcast channels where the optimal beamforming design was largely an open problem prior to the finding of the duality. Inspired by this finding, we address a similar question in the current work: \emph{What is the downlink dual of the (uplink) AirComp beamforming for sensor networks?} 

\begin{table}[!tt]
\centering
\caption{ Examples of nomographic functions (see Definition \ref{def:Nomographic}) that are AirComputable.}
\begin{tabular}{|p{4cm}|p{4cm}|}
\hline
\bf{Name} & \bf{Expression} \\
\hline
Arithmetic Mean &  $h = \frac{1}{K}\sum_{k=1}^K d_k$ \\ 
\hline
Weighted Sum &  $h = \sum_{k=1}^K \omega_k d_k$ \\ 
\hline
Geometric Mean &  $h = \l(\prod_{k=1}^K d_k \r)^{1/K}$ \\ 
\hline
Polynomial &  $h = \sum_{k=1}^K \omega_k d_k^{\beta_k}$ \\ 
\hline
Euclidean Norm &  $h = \sqrt{\sum_{k=1}^K d_k^2}$ \\ 
\hline
\end{tabular}
\label{summary:table1}
\end{table}

\subsection{Contributions and Organization}
We consider a multi-antenna multi-modal  sensor network where a multi-antenna  AP performing fusion of data transmitted by a cluster of multi-antenna multi-modal  sensors. By measuring multiple  time-varying parameters of the environment, each  sensor generates  multiple  data streams.  In each time slot,  a sensor transmits a set of multi-modal data values in the analog way,  namely by amplitude modulation \cite{GoldenbaumTCOM2013, GoldenbaumTSP2013}, over multiple antennas.  The transmissions of all sensors are simultaneous. The AP attempts to jointly receive multiple \emph{nomographic functions} (such as those in Table \ref{summary:table1}) of distributed sensor data by AirComp and spatial multiplexing. The AirComp of a  nomographic function is  implemented  by three cascaded operations: 1) pre-processing at sensors, 2) weighted summation of preprocessed outputs by simultaneous  transmissions, and 3) post-processing at the AP \cite{GoldenbaumTCOM2013, GoldenbaumTSP2013, KatabiAirComp2016}.  
In the current scenario, transmit and receive beamforming are applied to spatially multiplex multi-function AirComp as well as exploit spatial diversity to  minimize the distortion of function values caused by  noise, which is measured by sum \emph{mean-squared error} (MSE) over functions. 

While the traditional uni-function AirComp is a simple technique, the proposed multi-function version is challenging with the optimization of receive beamforming proved  to be NP-hard. Specifically, for uni-function AirComp, channel inversion at each sensor yields a desired weighted sum of preprocessed data at the AP, giving the desired function value after post-processing \cite{KatabiAirComp2016}. For multi-function AirComp, channel inversion remains optimal as shown in this work and is implemented by zero-forcing beamforming. Nevertheless, receive beamforming for multi-function AirComp,  referred to as \emph{MIMO-AirComp equalization},  is a new design problem  that finds no relevant result in the AirComp literature. The equalizer optimization is non-convex but can be relaxed as a \emph{semidefinite programming} (SDP) problem and thus solved using an iterative interior point algorithm. The standard approach does not yield any insight into the optimal equalizer structure and more importantly does not lead to an efficient channel-feedback design for acquiring the equalizer at the AP. To address these issues, we impose an  orthogonality constraint on the AirComp equalizer,  which is a technique for approximate beamformer design and limited channel feedback as widely applied  in the literature \cite{love2005limited,choi2006interpolation, peters2011cooperative, medra2015incremental}. Concretely, this allows a close-to-optimal equalizer to be derived in closed-form using  tools from differential-geometry, revealing an interesting  geometry structure in the design. Moreover, the closed-form solution leads to  an efficient channel feedback design that exploits the AirComp architecture for direct equalizer acquisition at the AP from  simultaneous feedback by all sensors.

The main contributions of this work are summarized as follows. 

\begin{itemize}
\item {\bf Multi-Function AirComp Beamforming}: As mentioned, while zero-forcing transmit beamforming is found to be optimal, the receive-beamformer optimization for sum-MSE minimization under transmission-power constraints    can be proved to be NP-hard. By tightening the constraints, the resultant approximate problem is found to involve optimization on a Grassmann manifold, which can be interpreted as the space of subspaces. This allows the application of differential geometry to solve the approximate problem. The derived solution shows the normalized receive beamformer to be the \emph{weighted centroid of a cluster of points on the manifold}, where each point represents the eigen-subspace of an individual MIMO channel and the corresponding weight its smallest eigenvalue. In addition, the optimal beamformer norm is also derived in closed-form.  Such a beamformer design allowing   efficient computation is verified by simulation to be close-to-optimal. 

\item {\bf AirComp-Multicasting Duality}: As a by-product of our investigation, for the special case of AirComp  with single-antenna (uni-modal) sensors, the problem of receive-beamforming optimization is discovered to have the identical form as the classic problem of multicast transmit-beamforming, thereby establishing a novel \emph{AirComp-multicasting duality}. The latter problem is known to be NP-hard and typically solved using the \emph{semidefinite relaxation} (SDR) method.  The significance of our  finding lies in allowing the solution method for AirComp beamforming to be transferable to multicast-beamforming. This yields  a new solution method  for the  latter with complexity much lower than  the existing SDR approach as the network scales up. 

\item {\bf  AirComp Channel Feedback}: Last, we solve the \emph{AirComp feedback} problem: How to efficiently acquire the derived AirComp beamformer at the AP, which depends on global \emph{channel-state information} (CSI), by sensor distributed transmissions based on \emph{local CSI}? Given channel reciprocity, it is discovered that the AirComp system architecture can be also used  for efficient feedback. The resultant number of feedback rounds is \emph{independent} of the sensor population, overcoming the drawback of traditional channel training. Novel feedback techniques are designed for sequential feedback of the normalized AirComp beamformer and beamformer norm  based on their  derived expressions, where each feedback round involves concurrent transmissions by all sensors. Essentially, the two techniques implement AirComp of two functions,  namely the weighted centroid of a set of matrices and the maximum of a set of scalar values. They are  hence  exclusively  for  multi-function  AirComp  and may not be applicable for traditional multiuser MIMO communication  systems where multiuser feedback  requires  orthogonal channels and  focuses on precoder  quantization \cite{love2008overview,huang2012stability}. 

\end{itemize}

The remainder of the paper is organized as follows. Section II introduces the AirComp system model. Section III presents the problem formulation for the enabling beamforming design and channel feedback. The proposed  beamforming design is presented in Section IV, and the duality between the uplink AirComp and downlink multicasting is discussed in Section V. Section VI proposes an efficient channel-feedback scheme that can be implemented by AirComp. Simulation results are provided in Section VII, followed by concluding remarks in Section VIII.

%

\begin{figure*}[tt]
\centering
  \includegraphics[width=0.72\textwidth]{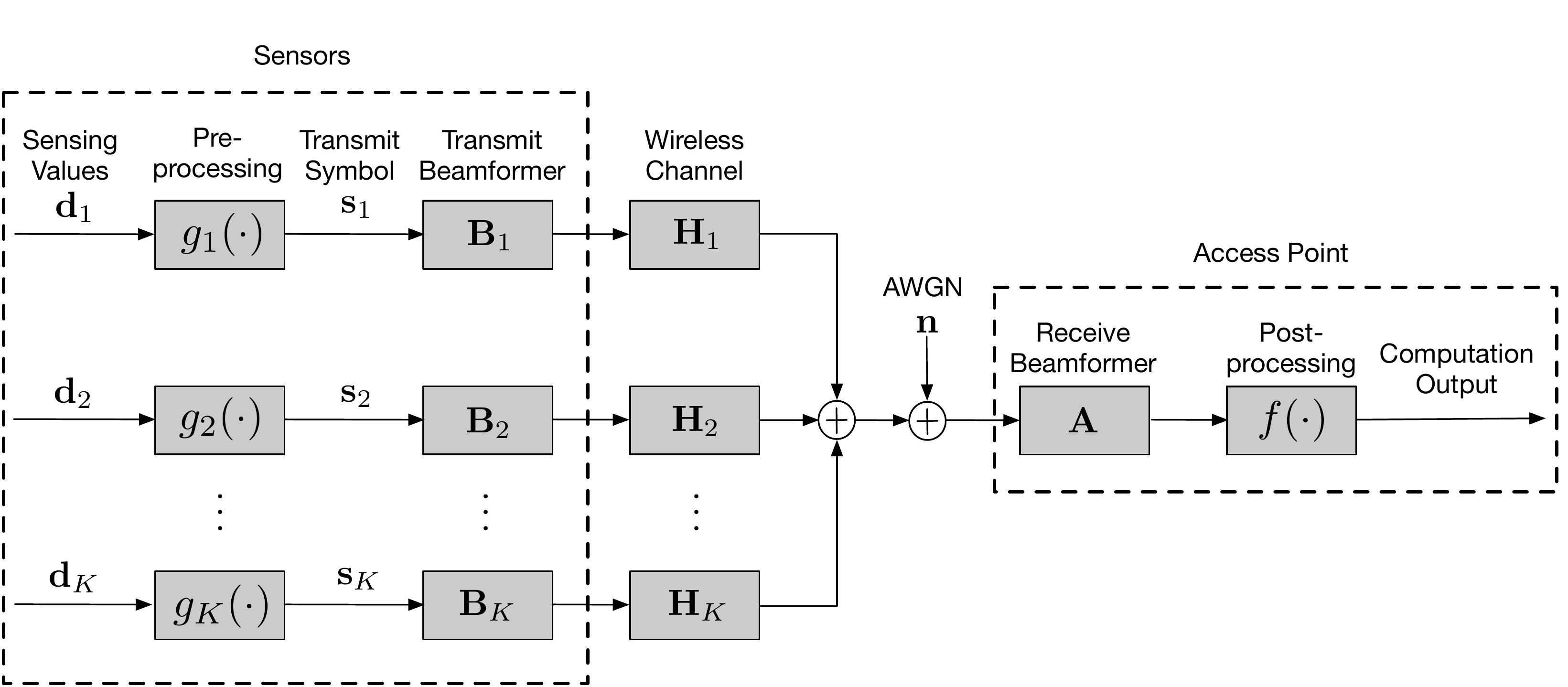}
\caption{System model of AirComp via MIMO transmission.}
\label{model2}
\vspace{-4mm}
\end{figure*}

\vspace{-3mm}
\section{System Model}
We consider a wireless sensor network consisting of $K$ multi-modal sensors and a single AP as illustrated in Fig.\ref{model2}. All nodes are equipped with  antenna arrays. Specifically, $N_t$ antennas are deployed at  each sensor and $N_r$ at the AP. Each multi-modal sensor records the values of  $L$ heterogeneous  time-varying parameters of the environment, e.g., temperature, pollution, and humidity. The data from the record of the $\ell$-th parameter is referred to as type-$\ell$ data. For an arbitrary time slot, the measurement vector of the $k$-th sensor, grouping   $L$ sample values,   is denoted by ${{\bf{d}}_k}= [ {{d_{k1}}, {d_{k2}},  \cdots,  {d_{kL}}} ]^T \in {\mathbb{R}^{L \times 1}}$, where $d_{k\ell}$ is the measurement of the parameter  $\ell$ at sensor $k$. Instead of collecting the whole data set, the AP aims at computing $L$ functions of $L$ corresponding types of data, denoted by $\{h_\ell(d_{1\ell},...,d_{K\ell})\}_{\ell =1}^L$, to support ultra-fast HMM sensing. The class of functions that are  computable by AirComp are called \emph{nomographic functions} as defined below.   

\begin{definition}[Nomographic Function \cite{GoldenbaumTSP2013}]\label{def:Nomographic}
\emph{The function $h_\ell(d_{1\ell},...,d_{K\ell})$ is said to be nomographic, if there exist $K$ preprocessing functions $g_{k\ell}(\cdot): \mathbb{R} \to \mathbb{R}$ along with a post-processing function $f_\ell(\cdot): \mathbb{R} \to \mathbb{R}$ such that it can be represented in the form:
\begin{align}\label{Nomographic Function}
h_\ell(d_{1\ell},...,d_{K\ell}) = f_\ell \l( \sum\nolimits_{k=1}^K g_{k\ell}(d_{k\ell}) \r). 
\end{align}
}
\end{definition}
\noindent Some common nomographic functions are listed in Table \ref{summary:table1}.  Based on the nested form of \eqref{Nomographic Function}, the AirComp of a nomographic function can be implemented in the sensor network by three operations as illustrated in Fig. \ref{model2}: 1) preprocessing at each sensor specified by $g_k =\{g_{k\ell}(\cdot)\}$ where $g_{k\ell}$ operates on the type-$\ell$ data at sensor $k$, 2) summation of preprocessed data realized by  multi-access, and 3) post-processing at the AP.  Considering  the computation of the geometric mean of type-$\ell$ data as an example,  the pre-processing  $g_{k\ell}(x) = \log x$, and the post-processing $f_\ell(y) = \exp(y/K)$. 
 Let $\bs_k =[g_{k1}(d_{k1}), g_{k2}(d_{k2}),  \cdots,  g_{kL}(d_{kL})]^T$ denote the multi-modal symbol vector after preprocessing and $\bh =[h_1, h_2, \cdots, h_L]^T$ the AirComputed function values.  For ease of transmission-power control and without loss of generality, the symbols are assumed to be normalized to have unit variance, i.e., $\mathsf{E}\{\bs_k \bs_k^H\} =  \bI$, where the normalization factor for each data type is uniform for all sensors and can be inverted at the AP. Given the one-to-one mapping between $\bs = \sum_{k=1}^K \bs_k$ and $\bh$ according to  \eqref{Nomographic Function}, we refer to $\bs$ as the \emph{target-function vector}   for ease of exposition. 

\subsection{AirComp Phase}
Assuming symbol-level synchronization\footnote{This can be achieved by broadcasting a common reference clock from the AP to sensors using the AirShare solution developed in \cite{KatabiAirshare2015}.}, all users transmit their symbol vectors simultaneously using their arrays.  The distortion of the received vector with respect to the target-function vector  due to MIMO channels and noise is suppressed using transmit and receive beamforming. In other words, the joint beamforming attempts to attain coherent combining  of $K$  symbol vectors  at the AP. Let  $\bA \in \mathbb{C}^{N_r\times L}$ denote the receive beamforming matrix and $\bB_k \in \mathbb{C}^{N_t\times L}$ the transmit beamforming matrix at sensor $k$. Then the  symbol vector received by the AP after receive beamforming is given as 
\begin{align}\label{estimated_s}
\hat \bs = \bA^H \sum_{k=1}^K \bH_k \bB_k \bs_k + \bA^H \bn,
\end{align}
where $\bH_k  \in \mathbb{C}^{N_r\times N_t}$ represents the MIMO channel  matrix for the link from  the  sensor $k$ to the  AP,  and $\bn$ is the \emph{additive white Gaussian noise} (AWGN) vector with  \emph{independent and identically distributed} (i.i.d.) $\mathcal{CN}(0, \sigma_n^2)$ elements. 
The distortion of $\hat \bs$ with respect to $\bs$, which quantifies the AirComp performance, is measured by the MSE defined as follows:
\begin{align}\label{def:MSE}
 \textsf{MSE}(\bf{\hat s}, \bs)= \E \l [\text{tr}\(({\bf{\hat s}}-\bs)({\bf{\hat s}}-\bs)^H\)\r ].
\end{align}

\noindent Substituting \eqref{estimated_s} into \eqref{def:MSE}, the MSE  can be explicitly written as a function of the transmit and receive beamformer as follows:
\begin{multline}\label{MSE_function}
 \!\! \textsf{MSE}(\bA, \{\bB_k\}) \!=\! \sum_{k=1}^K \text{tr}\( (\bA^H\bH_k \bB_k - \bI) (\bA^H\bH_k \bB_k - \bI)^H \) \\ + \sigma_n^2 \text{tr}\(\bA^H\bA\).
\end{multline}
The beamformers are optimized in the sequel under the criteria of MMSE.

\subsection{Channel Feedback Phase}
Consider the existence of channel reciprocity and assume that perfect local CSI is available at all sensors. One can infer  from \eqref{MSE_function} that computing the  MMSE receive beamformer requires global CSI, namely $\{\bH_k\}$. As mentioned, the naive approach of estimating the global CSI would incur long latency and large overhead when the number of sensors is large, thus is impractical for the ultra-fast HMM sensing applications. An intelligent channel-feedback design is proposed in the sequel to allow beamformer acquisition via  concurrent transmissions by all sensors.  Let $\bX_k \in \mathbb{C}^{N_t \times T}$ denote the signal matrix transmitted by sensor $k$  where $T$ is the signal length in symbol. Given typical high transmission power for channel training and feedback, the  feedback observation at the AP can be assumed to be noiseless and  thus be represented by an $N_r\times T$ matrix ${\bf Y}$ as follows:
\begin{align}(\text{AirComp Feedback})\label{feedback} \qquad
\bY = \sum_{k=1}^K \bH_k \bX_k .
\end{align}
As is clear in the sequel,  with proper design of $\{\bX_k\}$, $\bY$ can serve  as a sufficient surrogate of the global  CSI $\{\bH_k\}$ in  beamformer computation at the AP.

\section{Problem Formulation}

\subsection{AirComp Beamforming  Problem}
Consider  the joint optimization  of the transmit and receive beamformers under the MMSE criterion and the  transmission-power constraints. It is assumed that the average transmission power of each sensor cannot exceed a given positive value $P_0$. Since the transmitted symbols have unit variance,  the power constraints are given as 
\begin{align}\label{power_cons}
\|\bB_k\|^2 \leq P_0, \quad k = 1, 2, \cdots, K.
\end{align}
Following a common approach in the MIMO beamforming literature (see e.g., \cite{love2005limited,choi2006interpolation,peters2011cooperative,medra2015incremental}), the receive beamformer $\bA$ is constrained to be orthonormal matrix. As mentioned, the constraint can lead to a closed-form suboptimal solution with only marginal performance loss and furthermore facilitate efficient channel feedback design as presented in Section \ref{sec:feedback}. 
As pointed out in \cite{medra2015incremental}, for most communication objectives, it is the subspace spanned by the beamformer but not the exact beamformer  that has a crucial effect on the system performance, justifying  the said  constraint.
 Furthermore, under the MMSE criterion, a  positive scaling factor $\eta$,  called \emph{denoising factor}, is included in $\bA$ for regulating the  tradeoff between noise reduction and transmission-power control. To be specific,  reducing $\eta$ suppresses noise but requires larger transmission power to maintain the  MSE of computed function values. Mathematically, we can write $\bA = \sqrt{\eta} \bF$ with $\bF$ being a tall unitary matrix and thus $\bF^H\bF = \bI$. 
Then given the MSE in \eqref{MSE_function},  the MMSE beamforming problem can be  formulated as:

\begin{equation}({\bf P1})\qquad 
\begin{aligned}
\mathop {\min }\limits_{\eta, \bA, \{\bB_k\}} \; &  \textsf{MSE}(\bA, \{\bB_k\})  \\
{\textmd{s.t.}}\;\;&\|\bB_k\|^2 \leq P_0, \;\forall k,\\ 
& \bA^H \bA = \eta \bI. 
\end{aligned}
\nn
\end{equation}

The problem is solved in Section \ref{sec:beam}. 

\subsection{AirComp Channel  Feedback Problem}
We propose the use of the AirComp architecture in Fig. \ref{model2} to realize the mentioned  receive-beamformer acquisition by concurrent transmissions by all sensors.  Let $\bA^*$ denote the derived beamformer solution to problem P1 and $\tilde{f}$ and $\tilde{g}_k$ be the feedback counterparts of the AirComp operations $f$ and $g_k$  (see Fig. \ref{model2}). The key design constraint is that the transmitted signal $\bX_k$ in \eqref{feedback} must be a function of local CSI $\bH_k$ only, denoted as $\bX_k = \tilde{g}_k(\bH_k)$.  Then it follows that 
\begin{equation}({\bf P2})
 \qquad \bA^* = \tilde{f} \(\sum\nolimits_{k=1}^K \bH_k \tilde{g}_k (\bH_k) \).\nn
\end{equation}
and the problem of AirComp feedback design reduces to the design of the functions  $\tilde{f}$ and $\{\tilde{g}_k\}$. The solution is presented in  Section \ref{sec:feedback}.

\section{Multi-Function AirComp: Beamforming}\label{sec:beam}
In this section, the AirComp beamforming problem in Problem  P1 is solved. While zero-forcing transmit beamforming can be proved to be optimal, the receive beamforming optimization is found to be NP-hard. An approximate problem is obtained by tightening the power constraints. This problem allows a practical solution approach based on \emph{differential geometry}. The solution reveals that the optimal  receive beamformer can be approximated by the  weighted centroids of a cluster of points on a  Grassmann manifold, each corresponding to the subspace of an individual MIMO channel.  To facilitate exposition, some basic definitions and principles of Grassmann manifolds are provided in Appendix \ref{preliminaries:Grassmann}.
 
Problem P1 is difficult to solve due to its non-convexity. The lack of convexity arises from the coupling between transmit and receive beamformers in the objective function, and the orthogonality constraint on the receive beamformer. To simplify the problem, zero-forcing transmit beamforming conditioned on a receive beamformer is first shown to be optimal as follows. 

\begin{lemma}\label{ZF_precoding} \emph{
Given  a  receive beamformer $\bA$, the MSE objective stated in \eqref{MSE_function} is minimized by the following zero-forcing transmit beamformers:
\begin{align}\label{ZF}
\bB_k^* = (\bA^H \bH_k)^H (\bA^H \bH_k \bH_k^H \bA)^{-1}, \qquad \forall k.
\end{align}
}
\end{lemma}
\proof
See Appendix \ref{App:ZF_precoding}.
\endproof
We note that the power constraint imposed on the precoder ${\bB_k}$ will be enforced in the sequel via regulating the norm of the equalizer $\bA$, or equivalently, the denoising factor $\eta$.

\begin{remark}[Number of AirComputable  Functions]\emph{
Note that to ensure matrix $\bA^H \bH_k \bH_k^H \bA$ is invertible, it requires $L \leq \min\{N_t, N_r\}$. This implies that, the number of functions that can be simultaneously computed by the proposed multi-function  AirComp is limited by $\min\{N_t, N_r\}$. The result is due to the underpinning  limit of  MIMO spatial multiplexing: the maximum number of spatial streams is $\min\{N_t, N_r\}$.}
\end{remark}
 
 \vspace{-3mm}
By substituting \eqref{ZF} in Lemma \ref{ZF_precoding},  Problem P1 is  transformed to the equivalent problem of minimizing the denoising factor of the receive beamformer:
\begin{equation}({\bf P3})\qquad 
\begin{aligned}
\mathop {\min }\limits_{\eta, \bF} \; & {\eta}  \\
{\textmd{s.t.}}\;\;& \frac{1}{\eta} \text{tr} \((\bF^H \bH_k \bH_k^H \bF)^{-1}\) \leq P_0, \;\forall k,\\ 
& \bF^H \bF = \bI,
\end{aligned}
\nn
\end{equation}
where $\bF$ is defined earlier as  the normalized receive beamformer. Though Problem P3 has a simpler form than P1, it remains non-convex due to the non-convex orthonormal constraint of the receive beamformer. In fact, Problem P3 is found in the next section to be NP-hard via proving its  equivalence to the NP-hard multicast beamforming problem. To develop a tractable approximation of the problem, a reasonable modification of the power  constraints is derived. To this end, a useful inequality is obtained as shown below. 
\begin{lemma}\label{lemma:2}\emph{
Let $\bH_k = \bU_k \Sigma_k \bV_k^H$ denote the compact form of \emph{singular value decomposition}  (SVD) of $\bH_k$. Then  we have the following inequality:
\begin{align}\label{ineq:1}
\text{tr} \((\bF^H \bH_k \bH_k^H \bF)^{-1}\) \leq \frac{L}{\lambda_{\min}(\Sigma_k^2)\lambda_{\min}(\bU_k^H \bF \bF^H \bU_k) },
\end{align}
where the equality holds given a well-conditioned channel, i.e., $\Sigma_k = \lambda \bI$ for some constant $\lambda$.
\proof
See Appendix \ref{App:lemma:2}.
\endproof}
\end{lemma}
\noindent Tightening the power constraints in Problem P3 using Lemma~\ref{lemma:2} gives the approximate problem: 
\begin{equation}({\bf P4})\qquad 
\begin{aligned}
\mathop {\min }\limits_{\eta, \bF} \; & {\eta}  \\
{\textmd{s.t.}}\;\;&   \eta \lambda_{\min}(\bU_k^H \bF \bF^H \bU_k)\geq \frac{L}{P_0\lambda_{\min}(\Sigma_k^2)}, \;\forall k,\\ 
& \bF^H \bF = \bI.
\end{aligned}
\nn
\end{equation}
Since the feasible set of Problem P4 is smaller than that of P3, the solution to P4 is a feasible solution though potentially a suboptimal one to P3.  To solve Problem P4 using differential geometry, an equivalent form containing subspace distances between the  receive beamformer and individual MIMO channels is obtained as follows. 

\begin{lemma}\label{lemma:3}\emph{
The problem P4 and the following problem P5 are equivalent.
 \begin{equation}({\bf P5})\qquad 
\begin{aligned}
\mathop {\min }\limits_{ \bF} \;  \max_k & \;\; \lambda_{\min}(\Sigma_k^2) \l( d_{\sf P2}^2 (\bU_k, \bF) -1 \r) \\
{\textmd{s.t.}}\;\; & \bF^H \bF = \bI,
\end{aligned}
\nn
\end{equation}
where $d_{\sf P2}^2(\bU_k, \bF)$ denotes the projection 2-norm distance between the subspaces spanned by $\bU_k$ and $\bF$ [see \eqref{Projection 2-norm} in Appendix \ref{preliminaries:Grassmann}].
\proof
See Appendix \ref{App:lemma:3}.
\endproof
}
\end{lemma}
Problem P5  is not yet in a ready form admitting the differential-geometry solution approach and requires an additional approximation. For this purpose, the objective function is bounded below. 
\begin{lemma}\label{lemma:4}
\emph{The objective function in Problem P5 can be bounded as follows:
\begin{align}\label{objective_bounds}
&\frac{1}{K} \sum\nolimits_{k=1}^K \lambda_{\min}(\Sigma_k^2) d_{\sf P2}^2 (\bU_k, \bF) - \frac{c}{K} \notag\\
 \leq \; &\max_k  \;\lambda_{\min}(\Sigma_k^2) \l( d_{\sf P2}^2 (\bU_k, \bF) -1 \r)  \notag\\
 \leq  \;&\sum\nolimits_{k=1}^K \lambda_{\min}(\Sigma_k^2) d_{\sf P2}^2 (\bU_k, \bF) - c,
\end{align}
where we define $c = \sum_{k=1}^K \lambda_{\min}\l( \Sigma_k^2\r)$ which is a constant independent of the control variable $\bF$.
}
\end{lemma}
The proof is straightforward and omitted for brevity.  Approximating the objective function in P5 by either the lower or the upper bound in Lemma \ref{lemma:4} both lead to the same approximate problem which is given by:
 \begin{equation}({\bf P6})\qquad 
\begin{aligned}
\mathop {\min }\limits_{ \bF} \;  \sum_{k=1}^K & \;\; \lambda_{\min}(\Sigma_k^2) d_{\sf P2}^2 (\bU_k, \bF) \\
{\textmd{s.t.}}\;\; & \bF^H \bF = \bI. 
\end{aligned}
\nn
\end{equation}

\begin{remark}[Beamformer Geometric Interpretation]\emph{
 Problem P6 allows a geometric interpretation of the desired receive beamformer $\bF^*$. In fact, the problem is to find a \emph{weighted centroid} of a set 
of points (each being a subspace)   $\{\bU_k\}$ on a Grassmann manifold with the squared projection 2-norm as the distance  metric, where the weights are $\{\lambda_{\min}(\Sigma_k^2)\}$. This reveals that the receive beamformer makes the best-effort to be aligned  with all the $K$ MIMO channel matrices with the alignment (subspace) distances adjusted by corresponding  channel gains as specified by the smallest channel eigenvalues. 
} 
\end{remark}

Problem  P6 can be approximately solved by a closed-form  solution that can be efficiently computed without resorting to an  iterative algorithm. Particularly, the closed-form solution can be derived  by replacing the projection 2-norm distance $d_{\sf P2}$  with the projection F-norm $d_{\sf PF} (\bU_k, \bF)$ (see Appendix \ref{preliminaries:Grassmann}). Note that $d_{\sf P2} (\bU_k, \bF) \approx d_{\sf PF} (\bU_k, \bF)$ for a small principal angle \cite{edelman1998geometry} and are exactly equivalent  in the case of $N_t = 1$ according to  \eqref{dist_ineq:2}. Thus, the problem P6 can be approximated as 
 \begin{equation}\qquad 
\begin{aligned}
\mathop {\min }\limits_{ \bF} \;  \sum_{k=1}^K & \;\; \lambda_{\min}(\Sigma_k^2) d_{\sf PF}^2 (\bU_k, \bF) \\
{\textmd{s.t.}}\;\; & \bF^H \bF = \bI,
\end{aligned}
\label{P:6.1}
\end{equation}
which still seeks  a weighted centroid of channel subspaces as before but based on a different  subspace  distance metric.  
According to the definition in \eqref{Projection F-norm}, $d_{\sf PF}^2 (\bU_k, \bF)$ can be computed in a matrix form by 
\[d_{\sf PF}^2 (\bU_k, \bF) = N_t - \text{tr}(\bU_k^H \bF \bF^H \bU_k).\]
Then, substituting it to the objective function in Problem P7, the problem can be equivalently written as 
 \begin{equation}{(\bf P7)}\qquad 
\begin{aligned}
\mathop {\max }\limits_{ \bF} \;  \sum_{k=1}^K & \;\; \lambda_{\min}(\Sigma_k^2) \text{tr}(\bU_k^H \bF \bF^H \bU_k) \\
{\textmd{s.t.}}\;\; & \bF^H \bF = \bI.
\end{aligned}
\label{P:6.2}
\end{equation}
Problem  P7 remains  non-convex due to 1) maximimizing a convex objective function and 2) the orthogonality constraints on the variable $\bF$. Nevertheless, by intelligently constructing an equivalent unconstrained problem,  we are able to derive a closed-form solution for Problem  P7 (see the following Lemma \ref{lemma:5}) via analyzing the stationary points of the unconstrained problem. For ease of exposition, 
define a matrix   $\bG \in \mathbb{C}^{N_r \times N_r}$,  called \emph{effective CSI},  as follows:   
\begin{align}\label{CSI_function}
(\text{Effective CSI}) \qquad \bG = \sum_{k=1}^K \lambda_{\min}(\Sigma_k^2) \bU_k \bU_k^H.
\end{align}
As shown shortly in Lemma \ref{lemma:5},   the normalized AirComp receive beamformer $\bF^*$   depends on the global CSI via the effective CSI. In other words,  $\bG$ is sufficient for computing $\bF^*$. 

\begin{lemma} \label{lemma:5} \emph{
Let $\bG = \bV_{G} \Sigma_G \bV_{G}^H$ be the SVD of $\bG$ given in \eqref{CSI_function}. The solution of Problem  P7 is given by  the first $L$ principal eigen-vectors of $\bG$, namely
\begin{align}\label{Eq:RxBeam}
\bF^* = [\bV_G]_{:,1:L}.
\end{align}
\proof
See Appendix \ref{App:lemma:5}.
\endproof
}
\end{lemma}

Finally, combining Lemmas \ref{ZF_precoding} - \ref{lemma:5}, the proposed MMSE beamforming design for multi-function AirComp is  summarized as follows:
\begin{equation}\label{beamforming_design}
\boxed{
\begin{aligned}
\bullet \;\; &\text{Receive Beamformer}: \\ 
&\bA^* = \sqrt{\eta^*} \bF^*,  \quad \text{with} \ \bF^* \ \text{in \eqref{Eq:RxBeam}} ,\\
\bullet \;\; &\text{Denoising Factor}: \\ 
&\eta^* = {\max}_{k} \; \frac{1}{P_0} \text{tr} \(((\bF^*)^H \bH_k \bH_k^H \bF^*)^{-1}\), \\
\bullet \;\; &\text{Transmit Beamformer}: \\ 
&\bB_k^* = ((\bA^*)^H \bH_k)^H ((\bA^*)^H \bH_k \bH_k^H \bA^*)^{-1}, \qquad \forall k,
\end{aligned} 
}
\end{equation}
where the denoising factor $\eta^*$ is derived by examining the power constraints in Problem P3.

\section{AirComp-Multicasting Duality}\label{sec:duality}
\begin{figure*}[tt]
  \centering
  \subfigure[Downlink multicasting]{\label{subfig:dl_mc}\includegraphics[width=0.42\textwidth]{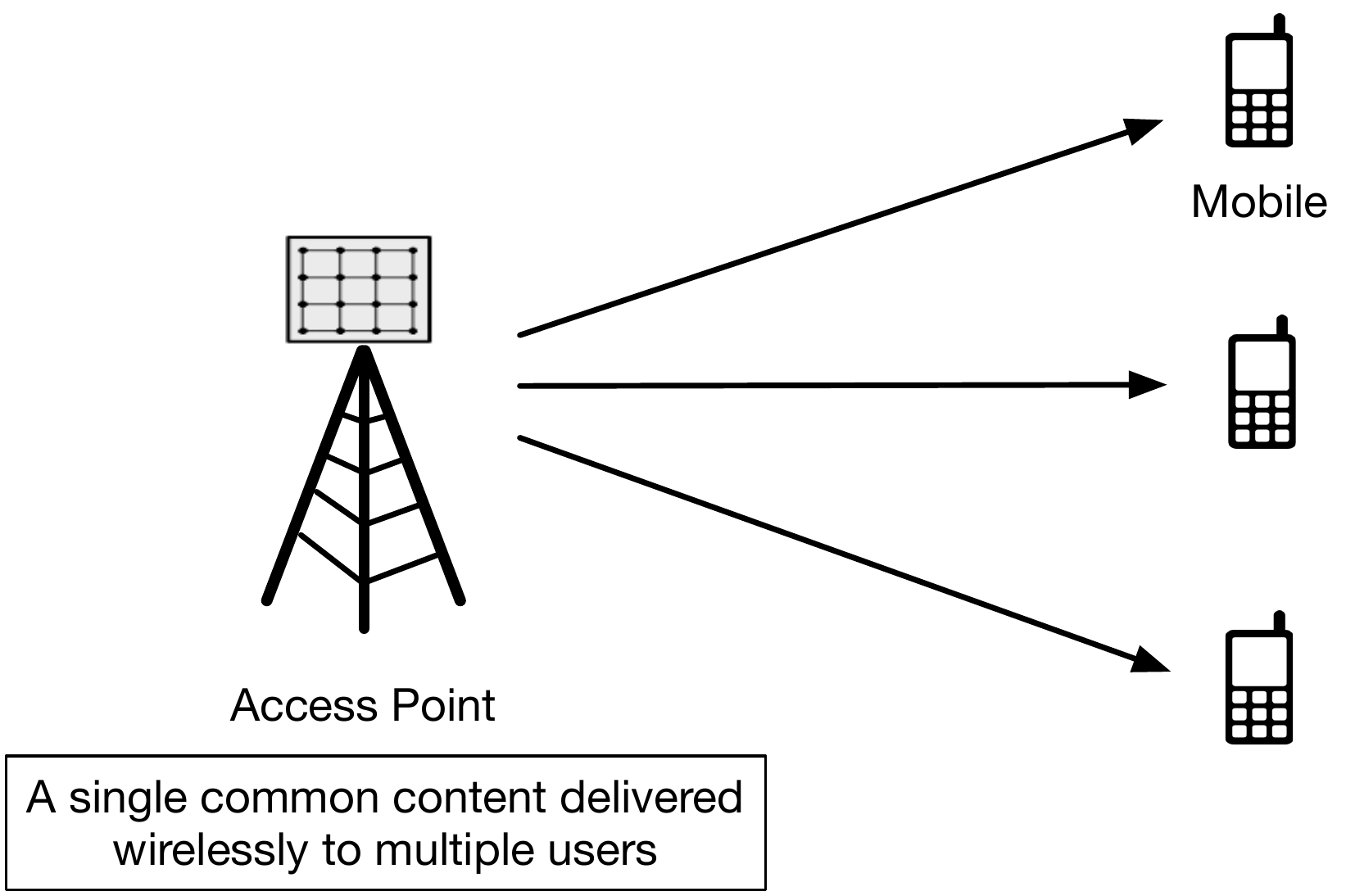}}
  \hspace{0.35in}
  \subfigure[Uplink AirComp]{\label{subfig:ul_ac}\includegraphics[width=0.42\textwidth]{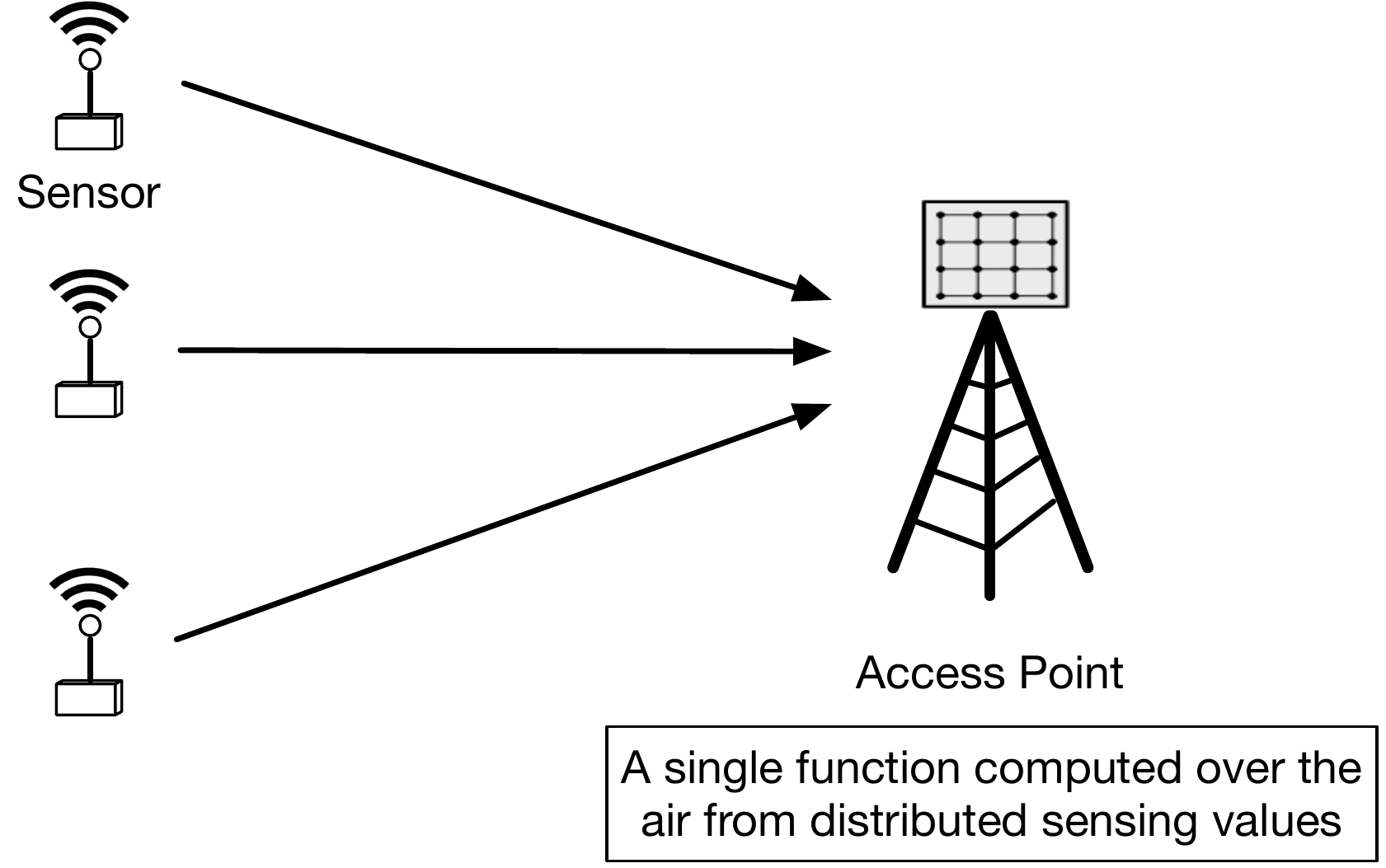}}
  \caption{The duality between downlink multicasting and uplink AirComp.}
  \label{Fig:dl_ul_duality}
  \vspace{-3mm}
\end{figure*}

\vspace{-1mm}
In this section, consider the case of single-antenna uni-modal sensors. The AirComp receive-beamforming problem for uplink sensing-data collection is shown to be equivalent to the well known problem of  transmit beamforming for downlink multicasting (see Fig.~\ref{Fig:dl_ul_duality}). This establishes the AirComp-multicasting duality, allowing the low-complexity beamforming design in the preceding section to be transferable to solve the NP-hard multicast beamforming problem.

\vspace{-2mm}
\subsection{Review of the Multicast Beamforming Problem}
\vspace{-1mm}
Consider  the scenario that multiple single-antenna users request the same data stream from a multi-antenna AP equipped  as shown in Fig. \ref{subfig:dl_mc}. Assuming global CSI  is available at AP, the problem of multicast beamforming is to minimize the total transmission  power subject to a set of \emph{signal-to-noise-ratio} (SNR)  constraints specifying the users' quality-of-service requirements. Mathematically, the problem can be formulated as follows:
\begin{equation}\qquad 
\begin{aligned}
\mathop {\min }\limits_{ \bw} \; & \|\bw\|^2 \\
{\textmd{s.t.}}\;\; & \frac{\|\bh_k^H \bw\|^2}{\sigma_k^2} \geq \gamma_k,\quad k = 1, 2, \cdots, K,
\end{aligned}
\label{multicasting_problem}
\end{equation}
where $\bw$ denotes the multicast beamforming  vector, $\bh_k$, $\sigma_k^2$ and $\gamma_k$ are the channel vector, noise variance and target SNR  of user $k$, respectively.
The problem can be  proved to be NP-hard \cite{sidiropoulos2006transmit}. Nevertheless, it is known that a close-to-optimal solution can be efficiently computed using the well known SDR technique. The key idea of SDR is to recast the problem as an equivalent rank-one constrained  SDP by denoting $\bW = \bw\bw^H$ as follows:
\begin{equation}\qquad 
\begin{aligned}
\mathop {\min }\limits_{ \bW} \; & \text{tr}(\bW) \\
{\textmd{s.t.}}\;\; & \frac{1}{\gamma_k \sigma_k^2} \text{tr} (\bh_k \bh_k^H \bW) \geq 1, \qquad \forall k,\\
& \bW \succeq \b0, \;\; \text{Rank}(\bW) = 1.
\end{aligned}
\label{SDR_problem}
\end{equation}
Then SDR drops the rank-one constraint and solves the relaxed SDP. Finally, a rank-one approximate solution of the original problem is retrieved by a Gaussian randomization strategy based on the solution of the relaxed SDP (or simply the principal eigenvector of it). For more details on the SDR algorithms, readers are referred to the key references in the area  \cite{sidiropoulos2006transmit,luo2010semidefinite}.

\subsection{Duality between  AirComp  and  Multicast  Beamforming}
For the special  case of a sensor network  with single-antenna sensors ($N_t =1$) and a multi-antenna AP. The receive beamforming reduces to a $N_r \times 1$ vector denoted by $\bf f$. Its  design problem for MMSE AirComp can be directly simplified from problem P5 (which is equivalent to the original P3 when $N_t =1$) to the following form:
 \begin{equation}\qquad 
\begin{aligned}
\mathop {\min }\limits_{ \bf f} \;  \max_k  \;\; &\|\bh_k\|^2 \l( d_{\sf P2}^2 (\bu_k, \bf f) -1 \r) \\
{\textmd{s.t.}}\;\; & \| {\bf f}\|^2 = 1,
\end{aligned}
\label{single_function_AirComp:1}
\end{equation}
where $\bu_k$ represents the orientation of the channel vector, i.e., $\bu_k = \frac{\bh_k}{\|\bh_k\|}$.

\begin{lemma}\label{lemma:6}\emph{
The solution to the problem \eqref{single_function_AirComp:1} is the same as that to the following problem \eqref{single_function_AirComp:2} up to a scaling factor.
\begin{equation}\qquad 
\begin{aligned}
\mathop {\min }\limits_{\bf f} \; & \|\bf f\|^2 \\
{\textmd{s.t.}}\;\; & \|\bh_k^H \bff\|^2 \geq 1,\quad k = 1, 2, \cdots, K. 
\end{aligned}
\label{single_function_AirComp:2}
\end{equation}
\proof
See Appendix \ref{App:lemma:6}.
\endproof
}
\end{lemma}

\begin{remark}[AirComp-Multicasting Duality]\emph{Comparing \eqref{multicasting_problem} and \eqref{single_function_AirComp:2} reveals  that the beamforming problems  for  the uplink AirComp and downlink multicasting share the same mathematical form. This establishes the AirComp-multicasting duality that is analogous to the famous uplink-downlink duality for multiuser MIMO communication \cite{jindal2004duality}.  Intuitively, the AirComp-multicasting duality is  a result of the fact that  both the AirComp  and multicast beamformers must make the same best effort 
to be aligned with multiple vector channels (see Fig.~\ref{Fig:dl_ul_duality}) though for different objectives: one is to minimize the distortion in the computed function value and the other  maximize the minimum SNR. }
\end{remark}

The AirComp beamforming technique designed in the last section is based on computing  the weighted centroid on the Grassmann manifold. The duality allows the technique to be applied to multicast beamforming. Compared with the classic SDR method discussed in the preceding subsection, the AirComp beamforming  has the following two main advantages. 

\begin{itemize}
\item {\bf (Efficient CSI Feedback)}: The AirComp beamforming solution requires only the \emph{effective CSI} in \eqref{CSI_function} and thus enables the efficient ``one-shot" channel feedback scheme presented in the next section. In contrast, the SDR solution requires global CSI $\{\bh_k\}$ and thus requires all users to feed back their local CSI. This results in excessive channel training overhead when the number of users is large.

\item {\bf (Low Computation Complexity)}: As shown in Lemma \ref{lemma:5}, the weighted centroid solution requires only one-shot computation and has a relatively low complexity of $O(N_r^2)$ arising from the principal eigenvector calculation. However, the SDR requires first solving a SDP of dimension $N_r$, by an iterative interior point method, resulting in the complexity of $O(\max\{N_r, K\}^4 N_r^{1/2} \log{(1/\epsilon)})$ where $\epsilon$ denotes the solution accuracy \cite{luo2010semidefinite}. The complexity becomes overwhelming when the numbers of the receive antennas and/or users are large. This makes the AirComp  solution  preferable in practice. The low-complexity advantage  of the proposed solution is also verified by simulation in Section \ref{sec:simulation}.
\end{itemize}

\section{Multi-Function AirComp: Channel Feedback}\label{sec:feedback}
In this section, the AirComp  feedback problem stated in Problem P2 is solved by feedback technique design. Using the optimization results in Section \ref{sec:beam}, two novel techniques are designed in the following subsections for sequential feedback of two components of the AirComp receive beamformer, namely the normalized beamformer and the denoising factor. Essentially, the two techniques realize the AirComp of two functions: 1) the weighted centroid of a set of matrices and 2) the maximum of a set of scalars, which can be thus implemented using the AirComp architecture in Fig. \ref{model2}. 

\subsection{Feedback of Normalized Beamformer} 
Assume that the  feedback channel is noiseless due to  high transmission power for channel feedback. The  AirComp feedback scheme for  acquisition of the normalized beamformer $\bF^*$ in \eqref{Eq:RxBeam}  is derived as follows.  As indicated by Lemma \ref{lemma:5}, the normalized beamformer $\bF^*$    can be directly computed from effective CSI matrix  $\bG$ in \eqref{CSI_function}. The key step for solving the   AirComp feedback problem   in Problem P2 is to enforce the equality $\bY = \bG$. Then given $\bG = \sum_{k=1}^K \lambda_{\min}(\Sigma_k^2) \bU_k \bU_k^H$ and Let $\bH_k = \bU_k \Sigma_k \bV_k^H$ denote the compact SVD of $\bH_k$, it is easy to verify using Problem P2 that designing the feedback signals $\{\bX_k^*\}$ as $\bX_k^* = \lambda_{\min}(\Sigma_k^2) \bV_k \Sigma_k^{-1} \bU_k^H$ gives the desired equality $\bY = \bG$.  Then the normalized receive beamformer can be computed as the principal eigenvectors of the received signal $\bY = \bG$.  The AirComp feedback design for acquiring $\bF^*$ is summarized as follows. 

\noindent \underline{\bf Normalized Beamformer Feedback:} 
\vspace{2mm}
\begin{equation}\label{normalized_beam_feed}
\boxed{
\begin{aligned}
\bullet \;\;&\text{Feedback Signal}: \\  &\bX_k^* = \tilde{g}_k(\bH_k) =  \lambda_{\min}(\Sigma_k^2) \bV_k \Sigma_k^{-1} \bU_k^H, 
\\
\bullet \;\; &\text{Received  Signal}: \\ &\bY = \sum\nolimits_{k=1}^K \bH_k \bX_k,\\
\bullet \;\; &\text{Feedback Post Processing}: \\ &\bF^* = \tilde{f}(\bY) = [\bU_Y]_{:,1:L},
\end{aligned}
}
\vspace{2mm}
\end{equation}
where  $[\bU_Y]_{:,1:L}$ denote the $L$ dominant left eigenvectors of  $\bY$. It is important to note that scaling  the feedback signal $\bX_k^*$ by a constant so as to meet a transmission power constraint has no effect on the received normalized beamformer $\bF^*$. 

The above AirComp feedback technique inherits the advantage of AirComp by turning interference from multiple access into useful signals for functional computing. In contrast with the traditional method of channel training, increasing the number of simultaneous sensors may even be beneficial to the computation accuracy via sensing-noise averaging.

\subsection{Feedback of Denoising Factor}
Following \eqref{beamforming_design}, the denoising factor $\eta^*$  is the maximum of a set of scalars $\{\eta_k\}$  called feedback values and defined as: 
\begin{equation}
\eta_k = \frac{1}{P_0} \text{tr} \(((\bF^*)^H \bH_k \bH_k^H \bF^*)^{-1}\).\label{Eq:FbVal}
\end{equation}
Since the maximum  is not a nomographic function, it is  not directly  AirComputable. However, an intelligent feedback technique is presented shortly that shows the possibility of   denoising-factor acquisition  by AirComp over a fixed number of $M$ feedback rounds. To begin with, it is assumed that the normalized beamformer $\mathbf{F}^*$ is acquired at the AP using the technique in the preceding sub-section and then broadcasts to all sensors. This allows each sensor to apply zero-forcing transmit beamforming for inverting the corresponding channel matrix. Specifically, the transmit beamformer  at sensor $k$ is given as $\bB_k = ((\bF^*)^H \bH_k)^H ((\bF^*)^H \bH_k \bH_k^H \bF^*)^{-1}$. Such beamforming creates a effective set of \emph{parallel MACs} such that the signal vectors transmitted by sensors are summed at the AP. In other words, with $\mathbf{e}_k$ denoting the $N_t\times 1$ signal vector for sensor $k$, the receive signal vector at the AP is $\sum_k \mathbf{e}_k$. Furthermore, we assume that the denoising factors lie in a fixed finite range  $[\eta_{\min}, \eta_{\max}]$.  Given the parallel MACs and the assumption, the algorithm for denoising-factor feedback with $M$ feedback rounds is described as follows. 

\vspace{3mm}
\noindent \underline{\bf Algorithm for Denoising-Factor Feedback:} 

\begin{itemize}
\item[1)] {(\bf Initialization)}: Set the feedback counter $n = 1$ and intialize the feedback-quantization range  $[\eta_{\min}^{(n)}, \eta_{\max}^{(n)}]$ with   $\eta_{\max}^{(1)} = \eta_{\max}$, and $\eta_{\min}^{(1)} = \eta_{\min}$.

\item[2)] {(\bf Feedback Quantization)}: A quantizer codebook with $N_t$ values, denoted by $\bQ = [q_1,q_2,$
$\cdots, q_{N_t}]$, with $q_1<q_2<\cdots<q_{N_t}$, is generated by uniformly partitioning  the range $\l[\eta_{\min}^{(n)}, \eta_{\max}^{(n)}\r]$. Thus, the maximum quantization error $\epsilon_{\max}$ is bounded by half of each partition interval denoted as $\Delta$: 
$\epsilon_{\max}\leq  \frac{\Delta}{2} =  \frac{\eta_{\max}^{(n)} - \eta_{\min}^{(n)}}{2 N_t}$. Quantizing the feedback value $\eta_k$ in \eqref{Eq:FbVal} at sensor $k$ gives the codebook index $m_k = \arg \min_m |\eta_k - q_m |$.

\item[3)] {(\bf Concurrent  Feedback)}: Each sensor transmits a signal vector comprising a single $1$ at the location specified by the corresponding codebook index and $0$'s at other locations. Specifically, the signal vector ${\bf e}_k$ for sensor $k$ is 
\begin{align}
{\bf e}_k = [0,\cdots,0,1,0,\cdots,0], \quad \text{with}\quad [{\bf e}_k]_{m_k} = 1.
\end{align}
Then all sensors transmit their signal vectors simultaneously over the said effective parallel MACs.  The AP finds   the largest index of a \emph{nonzero element} in the   received signal vector $\sum_k {\bf e}_k$, denoted as the $\ell_{\max}$. Thereby, using the codebook  $\bQ$, it can be inferred at the AP that the quantized value of the  denoising factor $\eta^*$ is $q_{\ell_{\max}}$ and the exact value lies in the range $\l[q_{\ell_{\max}}-\frac{\Delta}{2}, q_{\ell_{\max}}+\frac{\Delta}{2}\r]$.

\item[4)] {(\bf Refining Quantization Range)}: To improve the quantization resolution in the next feedback round, the AP refines the quantization range as 
\begin{align}
\eta_{\max}^{(n+1)} = q_{\ell_{\max}} + \frac{\Delta}{2} \qquad \text{and} \qquad \eta_{\min}^{(n+1)} = q_{\ell_{\max}} - \frac{\Delta}{2}.\nn
\end{align}

Next, increase the counter by setting $n = n+1$ and go back to 2) if $n < M$ or  otherwise stop the feedback process.
  
\end{itemize}

From the above algorithm, the key result  on the feedback accuracy follows. 
\begin{proposition}\label{prop:denoising_feed}
\emph{Given  $M$ feedback rounds, the AP receives a quantized version  of $\eta^*$, denoted as  $\hat{\eta}^*$, with the  quantization error $\epsilon = |\hat{\eta}^* -\eta^* |$  bounded as 
\begin{align}\label{quantization_error}
\epsilon \leq \frac{\eta_{\max} - \eta_{\min}}{2} N_t^{-M}.
\end{align}
}
\end{proposition}
Proposition \ref{prop:denoising_feed} implies that the feedback error reduces exponentially with the number of feedback rounds. Specifically, adding a feedback round improves the feedback resolution by $\log_2 N_t$ bit.  
Given a target resolution with the maximum  quantization error  $\epsilon_{\max}$, the required number of feedback rounds by the proposed feedback scheme is given by 
\begin{align}\label{number_rounds}
M = \l\lceil \frac{\log_2(\eta_{\max} - \eta_{\min}) - \log_2 2\epsilon_{\max}}{\log_2 N_t} \r\rceil.
\end{align}
As an  example, for some practical  settings of $\eta_{\max} = 100$, $\eta_{\min}=0$, $\epsilon = 10^{-4}$, $N_t = 8$, the required number of rounds (feedback slots) is  $M = 7$ according to \eqref{number_rounds}. This is much smaller  than the number of sensors in a dense network which determines the feedback rounds if the traditional method of channel training is adopted. 

\begin{remark}[Comparison with State-of-the-Art]\emph{The state-of-the-art algorithm for AirComp of a maximum function was proposed in \cite{KatabiAirComp2016} based on a different principle from the current design. In \cite{KatabiAirComp2016}, the maximum of a set of distributed feedback values is progressively computed  at the AP by sequential detection of the bits in the binary representation of the desired value via scheduling transmitting sensors by broadcasting a threshold. As the result, each feedback round increases the feedback resolution by a \emph{single  bit} and the  algorithm cannot be straighforwardly extended to exploit spatial multiplexing. In contrast, by exploiting spatial channels for implementing  uniform quantization, the proposed feedback algorithm achieves \emph{multi-bit}  resolution improvement for each feedback round as mentioned earlier. 
}
\end{remark}

\subsection{Comparison with Conventional Channel Training} 
For  conventional multiuser channel training,  sensors take turns to transmit pilot signals to AP for uplink channel training to avoid collision (see e.g., \cite{gesbert2007shifting,spencer2004introduction}). The pilot signal for each sensor should be a  $N_t \times N_t$ or larger matrix  for estimating a $N_r \times N_t$ channel matrix. Thus it takes  at least $T = K \times N_t$ symbol  slots  to complete the channel training process  for a network comprising  $K$ sensors. In contrast, the proposed  AirComp feedback technique for normalized beamformer feedback involves simultaneous transmissions of all $\{\bX_k\}$, each of  size $N_t \times N_r$, which thus requires  only $N_r$ symbol  slots. This together with the $M$ slots (typically $6-8$) for the feedback of the denoising factor yields the total feedback slots of $N_r + M$ independent of the  network size $K$. Consider a typical dense sensor network with $K=100$, $N_r = N_t = 8$ and $M=8$, it takes only $16$ slots for the proposed AirComp feedback scheme in contrast to $800$ slots required by the conventional channel training. Thus AirComp feedback achieves  $50$-time  of feedback overhead reduction in this example.

\begin{figure*}[tt]
  \centering
  \subfigure[Effect of Number of Computed Functions]{\label{Fig:Effect of number of functions}\includegraphics[width=0.46\textwidth]{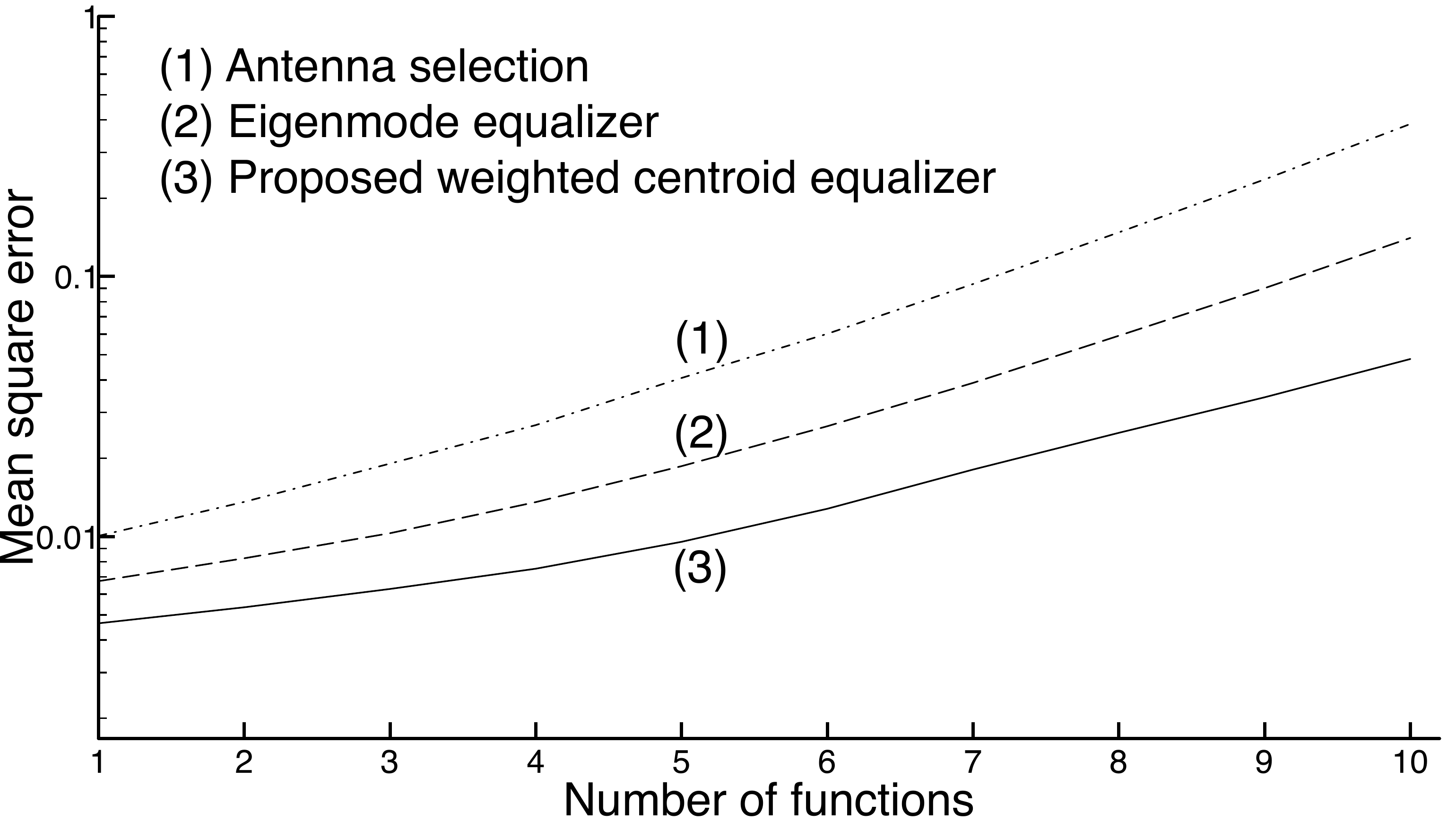}}
  \hspace{0.35in}  
  \subfigure[Effect of Receive Antenna Array Size]{\label{Fig:Effect of the antenna array size}\includegraphics[width=0.46\textwidth]{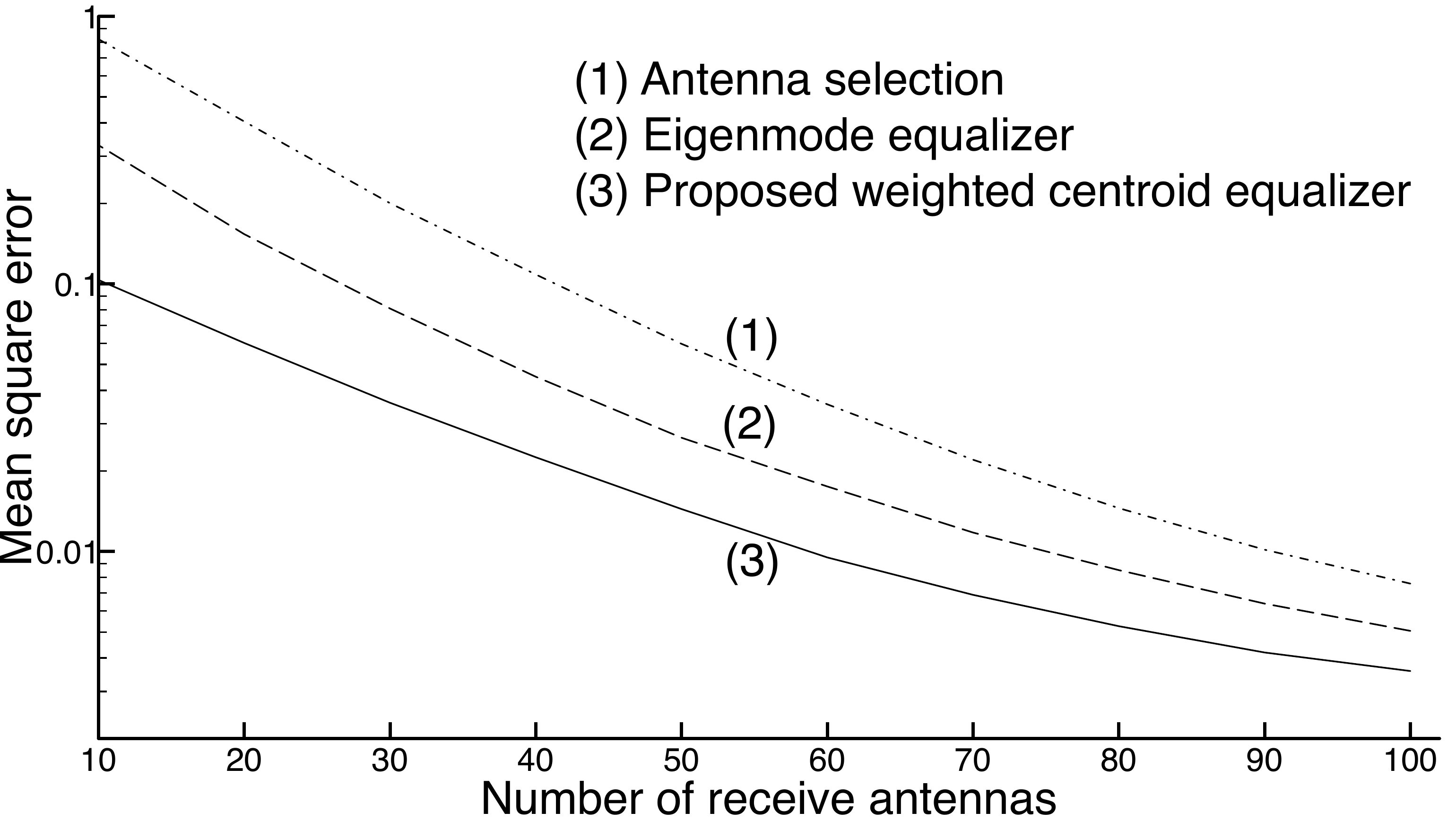}}
    \hspace{0.35in}  
  \subfigure[Effect of Network Scale]{\label{Fig:Effect of network scale}\includegraphics[width=0.46\textwidth]{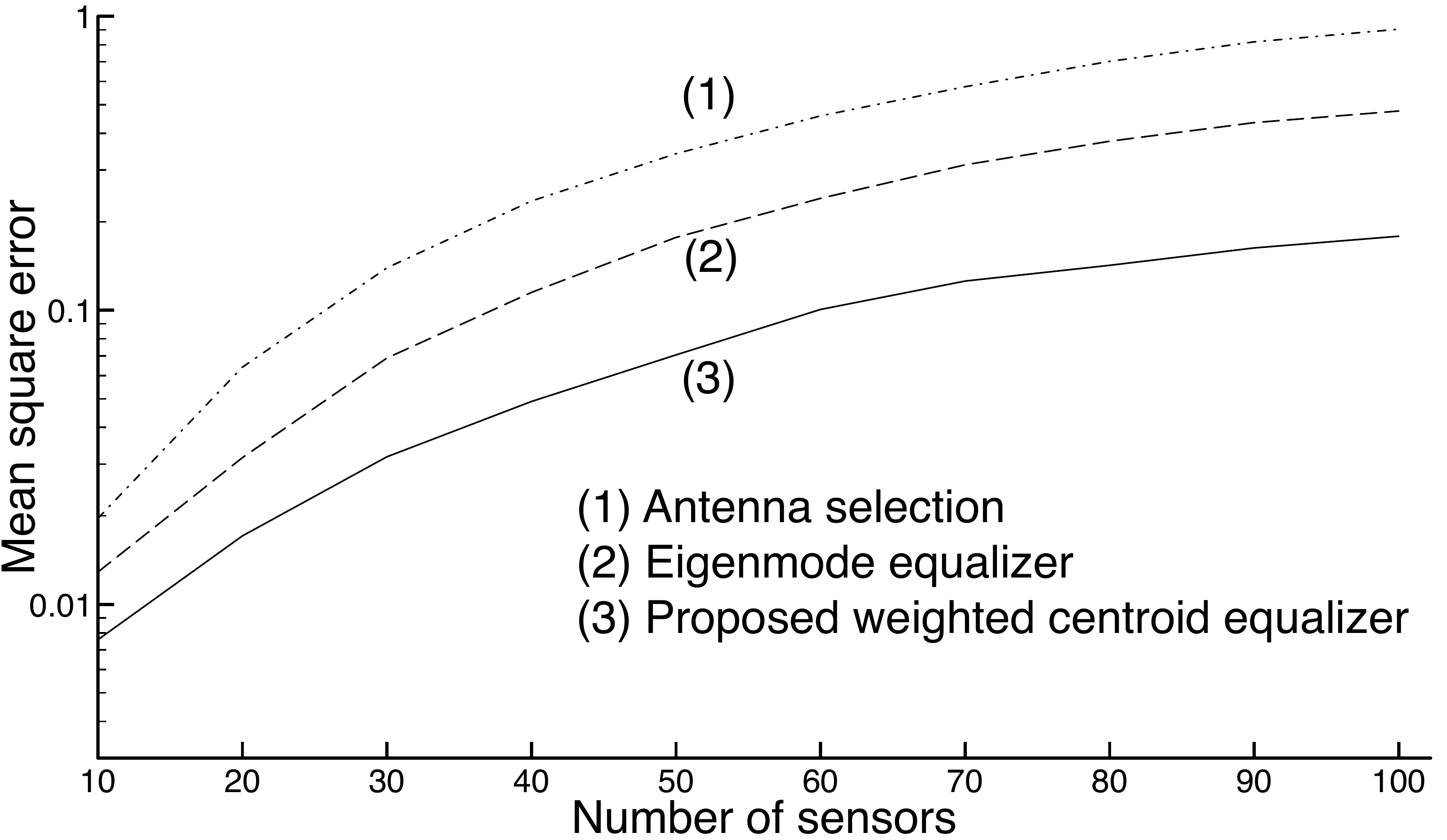}}
  \caption{Comparison between the proposed solution and the base line solutions for the case of multiple-function AirComp.}
  \label{Fig:Multi-function}
  \vspace{-4mm}
\end{figure*}

\section{Simulation Results}\label{sec:simulation}
In this section, the performance of the proposed multi-function AirComp is evaluated by simulation.  The simulation parameters are set as follows unless specified otherwise. The number of multi-modal sensors is  $K = 50$, the AP  array size at AP $N_r = 20$, the sensor array size  and the number of computed functions are equal to $N_t = L = 10$. Each MIMO  channel  are assumed to be i.i.d. \emph{Rician fading}, modelled as i.i.d. complex Gaussian random variables with non-zero mean $\mu = 1$ and variance $\sigma_h^2 = 1$.  In addition, the average transmit-SNR constraint, defined as $\rho_t = {P_0}/{\sigma_n^2}$, is set to be $10$ dB.

\subsection{Baseline Beamforming Schemes} 
Given that  the optimization of  AirComp beamforming is a NP-hard problem, for the purpose of comparison, we consider  two   baseline AirComp  beamforming schemes designed based on the classic approaches, namely \emph{antenna selection} and \emph{eigenmode beamforming}.  Both schemes assume zero-forcing transmit beamforming in \eqref{ZF} and their  difference lies in the receive beamformers.  Define the sum-channel matrix $\bH_{\sf sum} = \sum_{k=1}^K \bH_k$.  
To enhance the receive SNRs, the antenna-selection scheme selects the $L$ receive antennas observing the largest channel gains in the  sum channel   $\bH_{\sf sum}$. Consequently, the effective channel matrix after beamforming consists of  $L$ rows of $\bH_{\sf sum}$ with largest vector norms. On the other hand, to select the $L$ strongest eigenmodes of $\bH_{\sf sum}$ for AirComp, the  normalized eigenmode receive beamformer  consists of  the $L$ dominant left eigenvectors of  $\bH_{\sf sum}$. The denoising factor of each type of beamforming design is computed  following \eqref{beamforming_design} with $\mathbf{F}^*$ modified accordingly.

\begin{figure*}[tt]
  \centering
  \subfigure[MSE versus Receive Antenna Array Size]{\label{Fig:MSE vs N}\includegraphics[width=0.45\textwidth]{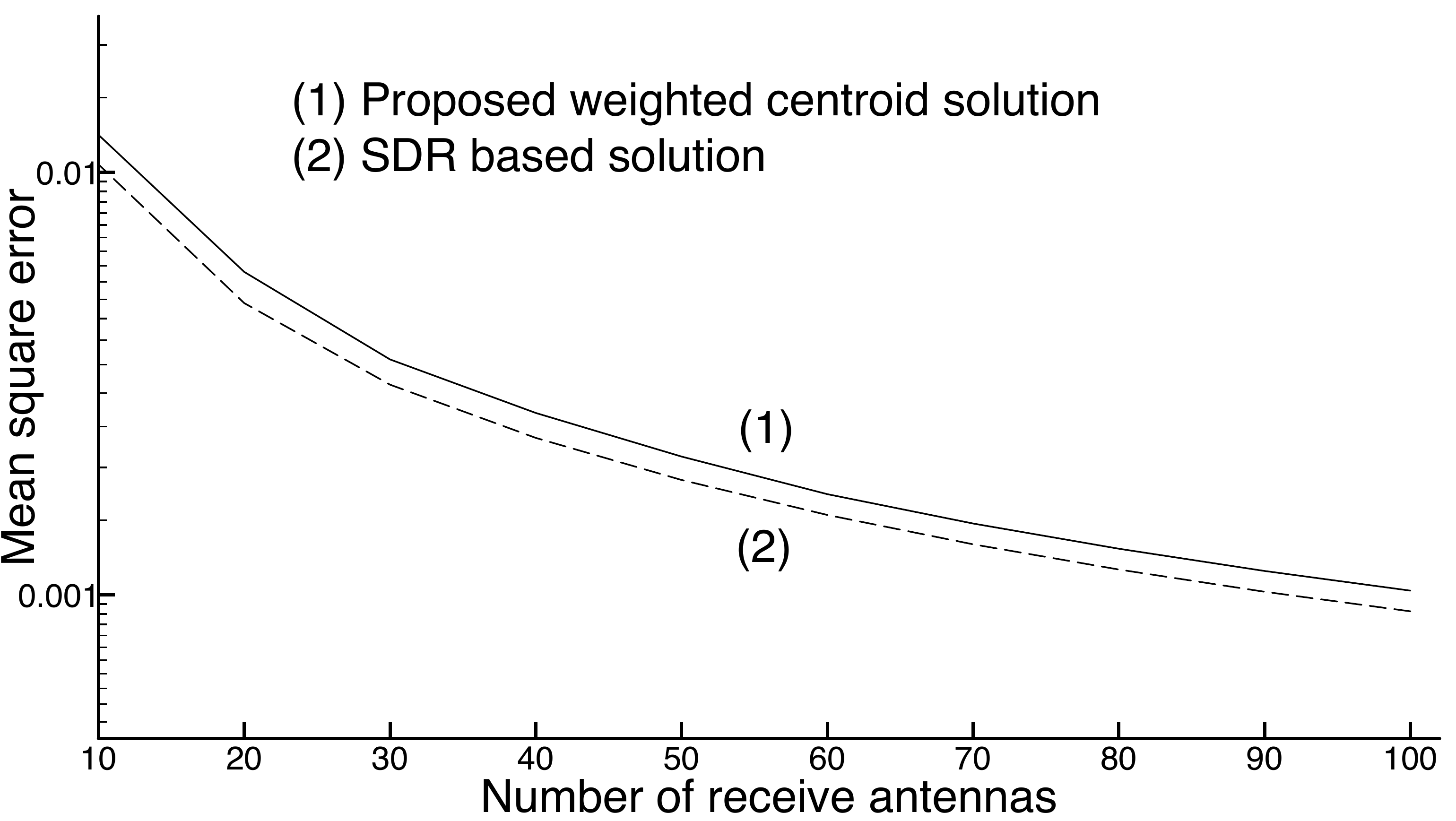}}
  \hspace{0.35in}  
  \subfigure[Execution Time versus Receive Antenna Array Size]{\label{Fig:time vs N}\includegraphics[width=0.45\textwidth]{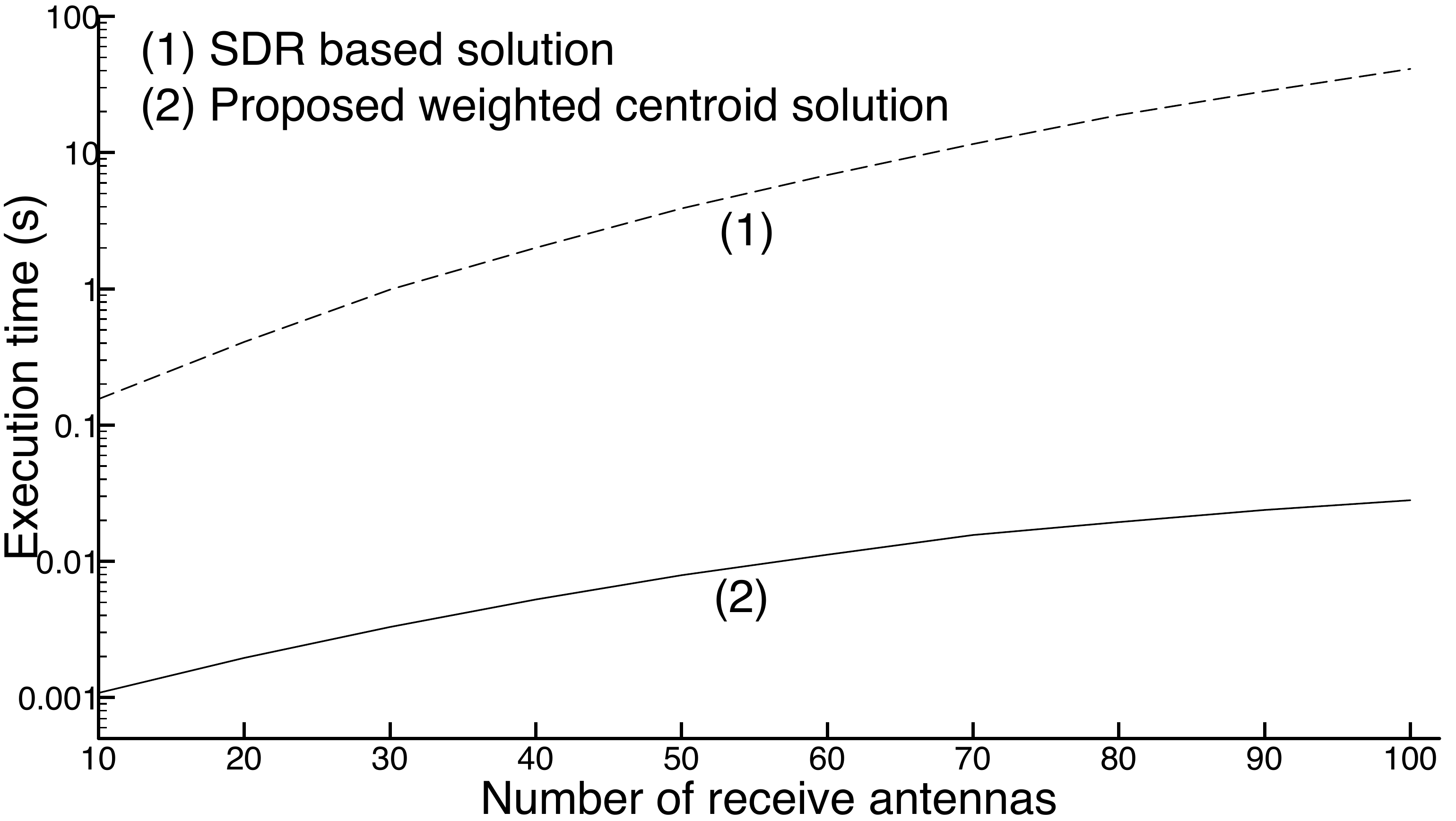}}
    \hspace{0.35in}  
  \subfigure[MSE versus Network Scale]{\label{Fig:MSE vs K}\includegraphics[width=0.45\textwidth]{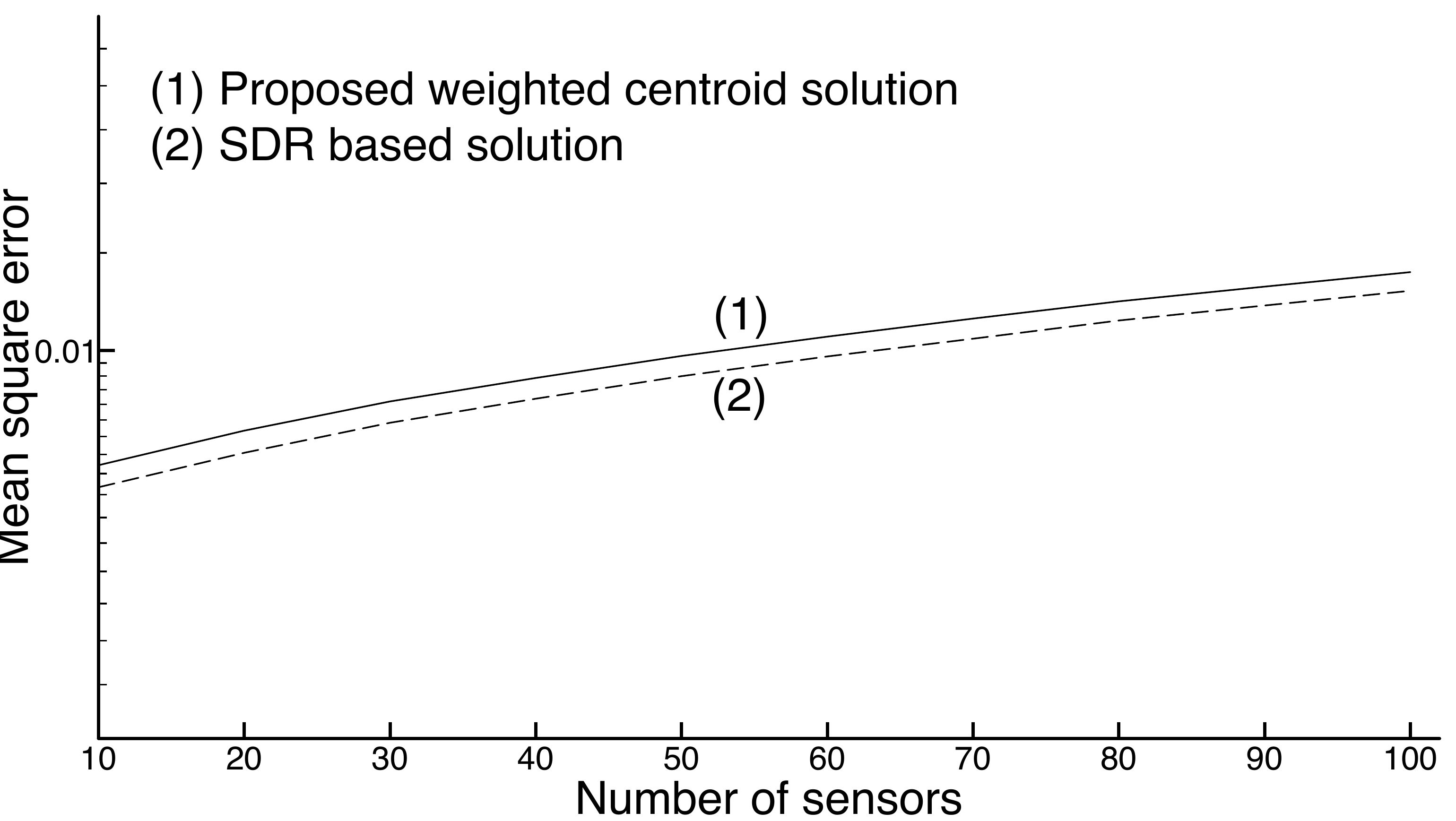}}
    \hspace{0.35in}  
  \subfigure[Execution Time versus Network Scale]{\label{Fig:time vs K}\includegraphics[width=0.45\textwidth]{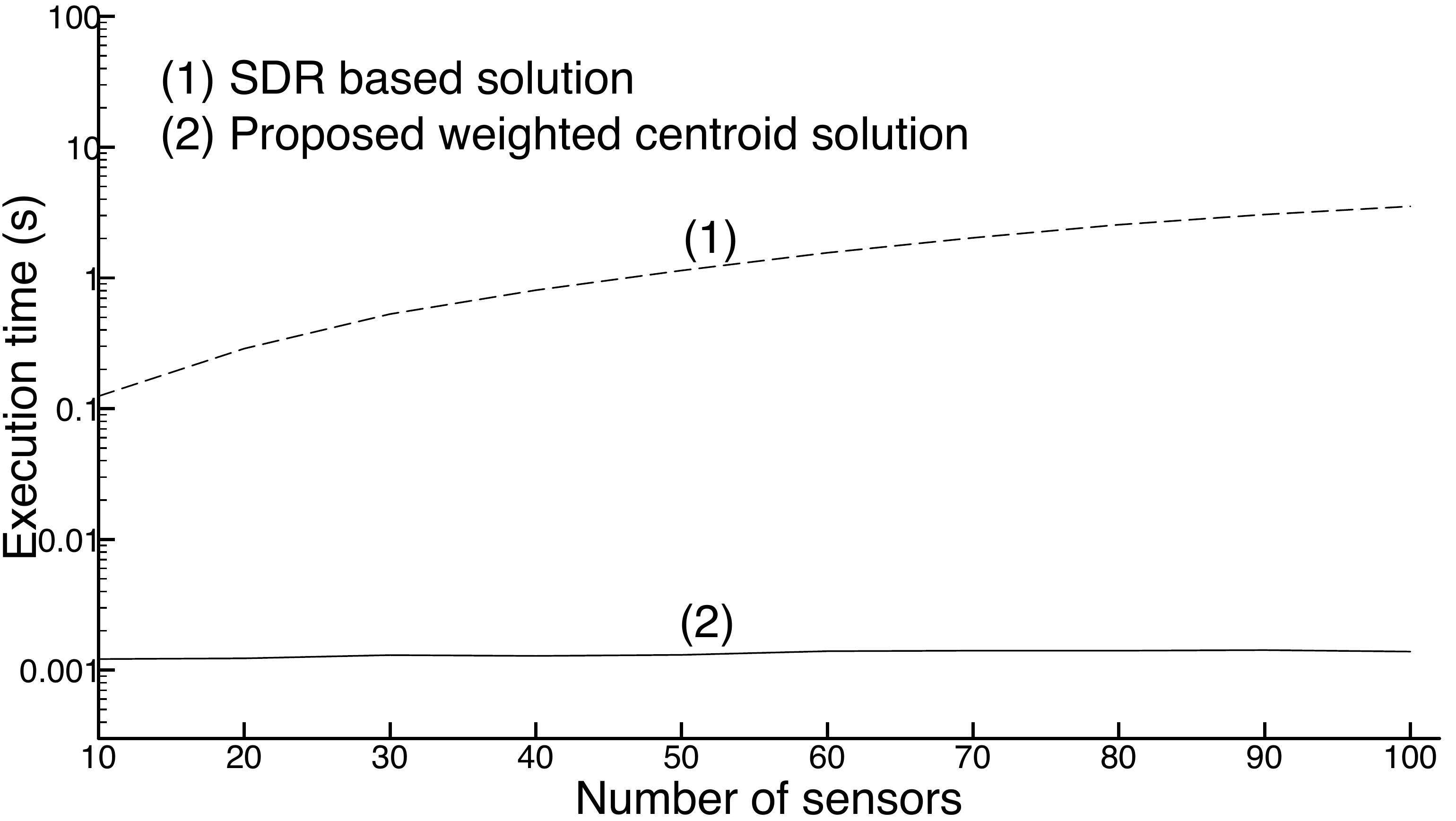}}
  \caption{Comparison between the proposed solution and the SDR based solution for the case of single-function AirComp.}
  \label{Fig:Single-function}
  \vspace{-4mm}
\end{figure*}

\subsection{Performance of Multiple-Function AirComp}

In Fig \ref{Fig:Multi-function}, the MSE performance of the proposed multi-function AirComp beamforming  is compared with that of  two baseline schemes introduced in the preceding subsections. A varying number of functions $L$, size of receive array $N_r$,  and also number of sensors $K$ are considered in Fig. \ref{Fig:Effect of number of functions} - \ref{Fig:Effect of network scale},  respectively. Several key observations can be made as follows. First, for all schemes, the MSE  is a increasing  function of $L$ and $K$ but  an increasing function of  $N_r$. This coincides with our intuition that, higher computation throughput is at a cost of declining  accuracy, and more connected sensors  makes it harder to design one common receive beamformer to equalize all different users' MIMO channels. Nevertheless, deploying more  receive antennas compensate for the performance degradation by exploiting  diversity gain. Second, under various parameter settings, the proposed scheme outperforms the other two baseline schemes, showing the effectiveness of the new design approach based on optimization on the Grassmann manifold. Furthermore, the performance gain of the proposed design  is larger  in the regime of large   $L$ and $K$, further confirming the effectiveness of the proposed design for multi-modal sensing and dense networks. Last, one can observe  that the performance  between different schemes converge as $N_r$ grows. This suggests that the large diversity gain enhances the receive SNRs such that the optimzation of AirComp beamforming is less critical and simple designs suffice.

\subsection{Comparison with the SDR Method}
The discovered AirComp-multicasting duality leads to the availability of two methods,   the proposed weighted centroid   and the SDR methods, for designing beamforming in either type of systems. Their performance and complexity are compared by simulation as follows. Consider single-antenna uni-modal sensors as in Section \ref{sec:duality}. 
The comparison of the MSE performance and computation time between the proposed weighted centroid and the SDR solutions is provided  in Fig. \ref{Fig:Single-function} for the varying  receive array size $N_r$ and number of sensors  $K$. The computation time is measured using  MATLAB. It is observed that the weighted centroid solution can achieve comparable performance as the SDR solution, which is optimal with a high probability  for the NP-hard multicast beamforming  problem as shown in \cite{luo2010semidefinite}. On the other hand, the former achieves dramatic computation time reduction with respect to the latter, ranging from 100x to 1000x in the considered ranges of $N_r$ and $K$. The simulation results support our previous analysis in Section \ref{sec:duality} that the complexity of the proposed solution is $O(N_r^2)$ independent of the network size  $K$ and  furthermore insensitive to the variation of the array size $N_r$ while the complexity  of the SDR solution is  $O(\max\{N_r, K\}^4 N_r^{1/2} \log{(1/\epsilon)})$. Thereby, the proposed solution features low complexity and is preferred in the large scale sensor networks (or large scale multicast  networks) or when the AP is equipped with a large scale array. 

\section{Concluding Remarks}
In this paper, we have proposed the framework of multi-function AirComp for MIMO HMM sensor networks. In particular, we have developed an approach for designing receive beamforming using tools from differential geometry. This approach achieves dramatic complexity reduction than the state-of-the-art SDR approach while maintaining comparable performance.  Furthermore, building on the AirComp system architecture, intelligent channel feedback techniques have been designed for enabling AirComp beamforming. Unlike the traditional method of channel training, the techniques prevent feedback overhead from escalating with the number of sensors and thus are highly efficient for dense HMM sensor networks. Last, the discovery of AirComp-multicasting duality allows the low-complexity beamforming design to be transferable to multi-antenna multicast systems, which traditionally relies on the computation-intensive SDR method for beamforming optimzation.  
The work points to the  promising new research area of MIMO AirComp where many interesting research issues  warrant further investigation such as sensor  scheduling, broadband AirComp, AirComp for multi-AP cooperative sensor networks, and AirComp for supporting distributed learning and inference.

\appendix
\subsection{Preliminaries on Grassmann Manifold}\label{preliminaries:Grassmann}
\subsubsection{Stiefel and Grassmann Manifolds}
The  $(N, M)$ Stiefel manifold is the set of all $N$-by-$M$ tall orthonormal matrices for $1\leq M \leq N$, denoted by ${\cal{V}}_{N,M}$. Mathematically, ${\cal{V}}_{N,M} = \{ {\bV}\in\mathbb{C}^{N\times{M}}:  {\bV}^{H} {\bV} = \bI \}$. On the other hand, the $(N, M)$  Grassmann manifold is a set of all ${M}$-dimensional subspaces in $\mathbb{C}^{N}$, denoted by ${\cal{G}}_{N,M}$. Thereby a Grassmann manifold ${\cal{G}}_{N,M}$ can be seen as the \emph{quotient space} of ${\cal{V}}_{N,M}$. To be specific, a point on the Grassmann manifold corresponds to  a class  of $N$-by-$M$ orthonormal matrices on the Stiefel manifold that span the same column subspace defined by the point. Choosing  an arbitrary matrix $\bU$ from this class and using  it as a \emph{generator}, the class, denoted as $[\bU]$, can be mathematically written as  $[\bU] = \{\bU \bQ: \bQ\in {\cal O}_{M} \}$ where ${\cal O}_{M}$ denotes  the group of  $M\times M$ unitary matrices. This leads to a   relation    between the Grassmannian ${\cal{G}}_{N,M}$ and the Stiefel  ${\cal{V}}_{N,M}$:  ${\cal{G}}_{N,M} = {\cal{V}}_{N,M}/{\cal O}_{M}$ \cite{edelman1998geometry}. 

\subsubsection{Distance Metrics on Grassmann manifold}
Algorithms on Grassmann manifold often involves the calculation of the distance between points  on the manifold. There exist   many different distance metrics and all of them  are derived  from a key notion called \emph{geodesic}. Roughly speaking, a geodesic is  the unique curve linking two points on a manifold that has  the shortest length among all. The length of the geodesic is called the geodesic distance (or arc length). Mathematically, given $\bX,\bY \in {\cal{G}}_{N,M}$, their geodesic distance, denoted as $d^2(\bX,\bY)$ is calculated by $d^2(\bX,\bY)  = \sum_{i=1}^M \theta_i^2$ where $\{{\theta}_i\}$ are called the \emph{principal angles}, measuring the minimal angles among any two sets of orthonormal bases spanning the two subspaces\cite{edelman1998geometry}. An efficient way to compute the principal angles is to perform SVD  on $\bX^H \bY$, i.e.,
\begin{equation}(\text{Principal Angles})\qquad  \label{principal_angle_SVD}
\bX^H \bY = \bU  \text{ diag}(\cos \theta_i) \bV^H,
\end{equation}
where the singular values in \eqref{principal_angle_SVD} yields the cosines of the principal angles. Based on these angles, a rich set of subspace-distance metrics can be defined. Two particular metrics of relevance in this paper are 
 \cite{edelman1998geometry}:
\begin{align}
\bullet \;\; &\text{Projection 2-norm}: \nn\\  &d_{\sf P2} (\bX, \bY) = \|\bX \bX^H - \bY \bY^H \|_2 = \| \sin (\boldsymbol\theta) \|_\infty, \label{Projection 2-norm}\\
\bullet \;\; &\text{Projection F-norm}: \nn\\  &d_{\sf PF} (\bX, \bY) = \|\bX \bX^H - \bY \bY^H \|_F = \|\sin (\boldsymbol\theta) \|_2, \label{Projection F-norm} 
\end{align}
where $\boldsymbol\theta$ is the vector formed by $\{\theta_i\}_{i=1}^M$, and matrices $\bU$ and $\bV$ follow those in \eqref{principal_angle_SVD}. For $\bX \neq \bY$, we have the following inequalities relating different distance metrics:
\begin{align}
\!\! d^2(\bX,\bY) \geq d_{\sf PF} (\bX, \bY) \geq d_{\sf P2} (\bX, \bY) \geq \frac{1}{\sqrt{M}} d_{\sf PF} (\bX, \bY). \!\!\label{dist_ineq:2}
\end{align}

\section{Supplementary Proof for the Derived Key Results}
\subsection{Proof of Lemma \ref{ZF_precoding}}\label{App:ZF_precoding}
Given the MSE objective provided in \eqref{MSE_function}, it is easy to note that both the first and the second terms within, i.e., $\sum_{k=1}^K \( (\bA^H\bH_k \bB_k - \bI) (\bA^H\bH_k \bB_k - \bI)^H \)$ and $\(\bA^H\bA\)$ are positive semidefinite  matrix with non-negative eigenvalues. As a result, for any given equalizer $\bA$, we have the following inequality:
\begin{multline}\label{app:psd_ineq}
 \sum_{k=1}^K \text{tr}\( (\bA^H\bH_k \bB_k - \bI) (\bA^H\bH_k \bB_k - \bI)^H \)\\ 
 + \sigma_n^2 \text{tr}\(\bA^H\bA\) \geq \sigma_n^2 \text{tr}\(\bA^H\bA\).
\end{multline}
It is easy to verify that setting $\{\bB_k\}$ to have the zero-forcing structure in \eqref{ZF} enforces  
\[\sum_{k=1}^K \text{tr} \( (\bA^H\bH_k \bB_k - \bI) (\bA^H\bH_k \bB_k - \bI)^H \) = 0,\]
and thus achieves the equality in \eqref{app:psd_ineq}, which completes the proof.

\subsection{Proof of Lemma \ref{lemma:2}}\label{App:lemma:2}
Utilizing the fact that, for any $n \times n$ square matrix, the inequality $\text{tr} ({\bf D}^{-1}) \leq {n}/{\lambda_{\min}({\bf D})}$ holds, it is straightforward to show 
\begin{align}\label{app:lemma2:1}
\text{tr} \((\bF^H \bH_k \bH_k^H \bF)^{-1}\) \leq \frac{L}{\lambda_{\min} (\bF^H \bH_k \bH_k^H \bF)}.
\end{align}
Note that $\bF$ is a tall unitary matrix, thus matrix $ (\bF^H \bH_k \bH_k^H \bF)$ and $\bH_k \bH_k^H$ share the same eigen-spectrum due to the well-known unitary invariant property. Therefore, it is easy to see that the upper bound \eqref{app:lemma2:1} becomes exact when the channel is well-conditioned.

Then given the compact eigenvalue decomposition $\bH_k \bH_k^H = \bU_k \Sigma_k^2 \bU_k^H$, we have the following
\begin{align}
 \lambda_{\min} (\bF^H \bH_k \bH_k^H \bF) &=   \lambda_{\min} (\bF^H \bU_k \Sigma_k^2 \bU_k^H \bF) \\
  & = \lambda_{\min} (\Sigma_k^2 \bU_k^H\bF \bF^H  \bU_k  ) \\
  & \geq \lambda_{\min} (\Sigma_k^2) \lambda_{\min}(\bU_k^H\bF \bF^H  \bU_k ) \label{app:lemma2:2},
\end{align}
where the second equality uses the fact that, for arbitrary two matrices $\bA$ and $\bB$ of the same size, $\bA \bB$ and $\bB \bA$ have the same eigen-spectrum, while the last inequality is due to \cite[Corollary 11]{zhang2006eigenvalue}, and it is easy to verify that the equality holds when $\Sigma_k^2$ is a scaled identity matrix, namely $\Sigma_k = \lambda \bI$  which also implies the channel is well-conditioned.

Finally, combining \eqref{app:lemma2:2} and \eqref{app:lemma2:1} leads to the desired result, which completes the proof.

\subsection{Proof of Lemma \ref{lemma:3}}\label{App:lemma:3}
Note that the set of power constraints in P4 can be rewritten as one single constraint by: 
\begin{align}\label{app:lemma3:1}
\eta \geq \max_k \frac{L}{P_0} \frac{1}{\lambda_{\min}(\Sigma_k^2)\lambda_{\min}(\bU_k^H \bF \bF^H \bU_k)}.
\end{align}
It is easy to note that the minimum $\eta$ in P4 is achieved when the above constraint is active (the equality holds). Therefore, one can move the constraint in \eqref{app:lemma3:1} to the objective function and have the following equivalent \emph{min-max} problem:
 \begin{equation}\qquad 
\begin{aligned}\label{Problem:5.1}
\mathop {\min }\limits_{ \bF} \;  \max_k & \frac{L}{P_0} \frac{1}{\lambda_{\min}(\Sigma_k^2)\lambda_{\min}(\bU_k^H \bF \bF^H \bU_k)} \\
{\textmd{s.t.}}\;\; & \bF^H \bF = \bI.
\end{aligned}
\end{equation}
The problem in \eqref{Problem:5.1} can be further simplified by \emph{max-min} the inverse of the objective function and dropping the constant term $\frac{L}{P_0}$ which leads to the following form
 \begin{equation}\label{Problem:5.2}\qquad 
\begin{aligned}
\mathop {\max }\limits_{ \bF} \;  \min_k & \;\; {\lambda_{\min}(\Sigma_k^2)\lambda_{\min}(\bU_k^H \bF \bF^H \bU_k)} \\
{\textmd{s.t.}}\;\; & \bF^H \bF = \bI.
\end{aligned}
\end{equation}
The objective function in \eqref{Problem:5.2} is related to the projection 2-norm Grassmannian metric via
\begin{align}\label{app:p2norm:def}
d_{\sf P2}^2 (\bU_k, \bF) = 1 - \lambda_{\min}(\bU_k^H \bF \bF^H \bU_k).
\end{align}
The equation can be easily verified by using the definitions provided in \eqref{principal_angle_SVD} and \eqref{Projection 2-norm}.

Finally, substituting \eqref{app:p2norm:def} into \eqref{Problem:5.2} leads to the desired problem P5, which completes the proof.

\subsection{Proof of Lemma \ref{lemma:5}}\label{App:lemma:5}
In this proof, we first construct an equivalent problem of P7 by modifying the objective function leveraging the orthogonality constraint, giving the problem ${(\bf P'')}$. Then, we show that solving a relaxed version of ${(\bf P'')}$ without the orthogonality constraint, denoted by ${(\bf P''')}$, yields a solution enforcing the constraint still. Therefore, one can conclude that problem P7 and ${(\bf P''')}$ share the same optimal solution, while the unconstrained problem ${(\bf P''')}$ can be solved easily by checking all stationary points of the objective. The detailed derivation is presented below. 

After some simple algebra manipulation exploiting the linearity of the trace operation, the objective function in (P7) can be further simplified as 
\begin{align}
\sum_{k=1}^K & \;\; \lambda_{\min}(\Sigma_k^2) \text{tr}(\bU_k^H \bF \bF^H \bU_k) = \text{tr}(\bF^H \bG \bF),
\end{align}
where $\bG$ is the effective CSI  that has been defined in \eqref{CSI_function}. Then problem (P7) reduces to 
 \begin{equation}({\bf P}') \qquad 
\begin{aligned}
\mathop {\max }\limits_{ \bF} \;  &\text{tr}(\bF^H \bG \bF) \\
{\textmd{s.t.}}\;\; & \bF^H \bF = \bI.
\end{aligned}
\label{P:6.3}
\end{equation}

Starting with problem ${(\bf P')}$, let's first define an alternative objective function $J(\bF)$ given by
\begin{align}
J(\bF) = -2 \text{tr}(\bF^H \bG \bF) + \text{tr} (\bF^H \bG \bF \bF^H \bF).
\end{align}
It is straightforward to verify that the following problem ${(\bf P'')}$ is equivalent to problem ${(\bf P')}$.
 \begin{equation}({\bf P''}) \qquad 
\begin{aligned}
\mathop {\min }\limits_{ \bF} \;  &J(\bF) \\
{\textmd{s.t.}}\;\; & \bF^H \bF = \bI.
\end{aligned}
\label{P:6.4}
\end{equation}
Now, let's relax the orthogonality constraint, and consider the following unconstrained problem:
 \begin{equation}({\bf P'''}) \qquad 
\begin{aligned}
\mathop {\min }\limits_{ \bF} \;  &J(\bF). \\
\end{aligned}
\label{P:6.5}
\end{equation}
Since the function $J(\bF)$ is a smooth function with gradient defined everywhere, the solution to the problem $({\bf P'''})$ should be a stationary point of $J(\bF)$, i.e., $\nabla J(\bF) = \b0$. It follows that
\begin{align}\label{grad_J:1}
\nabla J(\bF) = (-2 \bG + \bG \bF \bF^H + \bF \bF^H \bG) \bF = \b0.
\end{align}
From \eqref{grad_J:1}, one can note that $\bF = \b0$ is one of the stationary points. 

To seek other stationary points that $\bF \neq \b0$, we left multiply both sides of the equality in \eqref{grad_J:1} with $\bF^H$, which gives
\begin{align}\label{grad_J:2}
\bF^H \nabla J(\bF) = \bF^H \bG \bF (\bF^H \bF - \bI) +  (\bF^H \bF - \bI) \bF^H \bG \bF  = \b0.
\end{align}
Note that $\bF^H \bG \bF$ and $\bF^H \bF - \bI$ are symmetric matrices and  $\bF^H \bG \bF$ is positive definite, thereby one can conclude form \eqref{grad_J:2} that $\bF^H \bF = \bI$ by invoking the following lemma \cite{lutkepohlhandbook}:
\begin{lemma}\emph{
\!\!For hermitian matrices \!$\bX$\! and \!$\bY$\!, $\bX\bY \!+\! \bY\bX \!=\! \b0$ implies $\bY \!=\! \b0$ if $\bX$ is positive definite.  }
\end{lemma}
Next, substituting $\bF^H \bF = \bI$ into \eqref{grad_J:1} one can obtain the following equalities that a non-trivial stationary point ($\bF \neq \b0$) should enforce. 
\begin{align}\label{stationary_point}
\bG \bF = \bF \bF^H \bG \bF  \;\; \text{and} \;\;  \bF^H \bF = \bI.
\end{align}
It is easy to verify that stationary points satisfying \eqref{stationary_point} can be represented by $\bF = \bV_L \bQ$, where $\bV_L \in \mathbb{C}^{N_r \times L}$ collects $L$ distinct eigenvectors of $\bG$, and $\bQ$ is an arbitrary square unitary matrix. 

To find the global minimizer of $({\bf P'''})$, we evaluate $J(\bF)$ at each stationary point as follows:
\begin{itemize}
\item For $\bF = \b0$, we have $J(\bF) = 0$. 
\item For $\bF = \bV_L \bQ$, we have $J(\bF) = -\text{tr} (\bQ^H \bV_L^H \bG \bV_L \bQ) = -\text{tr} (\Sigma_L)$. 
\end{itemize}
where $\Sigma_L$ is a $L\times L$ diagonal matrix collecting the eigenvalues corresponding to the eigenvectors selected in $\bV_L$. Since $\bG$ is positive definite (see \eqref{CSI_function}), it follows that $-\text{tr} (\Sigma_L) < 0$, suggesting $\bF = \b0$ is not the global minimizer. Alternatively, one can see that the minimum of $J(\bF)$ is achieved by setting $\bF^* = \bV_L^* \bQ$ with $\bV_L^*$ being the $L$ principal eigenvectors of $\bG$. Note that the solution still enforces the orthogonality constraint, and thus also solves problem $({\bf P'})$.


\subsection{Proof of Lemma \ref{lemma:6}}\label{App:lemma:6}
Using the definition of the projection 2-norm in \eqref{Projection 2-norm} together with the fact that $d_{\sf P2} = d_{\sf PF}$ when $N_t = 1$, problem in \eqref{single_function_AirComp:1} can be rewritten as 
 \begin{equation}\label{Problem:lemma:6:1}\qquad 
\begin{aligned}
\mathop {\max }\limits_{ \bf f} \;  \min_k & \;\; \|\bh_k^H {\bf f} \|^2 \\
{\textmd{s.t.}}\;\; & \|\bf f\|^2 = 1.
\end{aligned}
\end{equation}
Then, by introducing a auxiliary variable $\rho = \min_k\; \|\bh_k^H {\bf f} \|^2$, the above problem can be equivalently reformulated as follows:
\begin{equation}
\begin{aligned}
\mathop{\min}\limits_{\bf f, \rho} \;  & \frac{1}{\rho} \|{\bf f}\|^2 \\
{\textmd{s.t.}}\;\; & \rho \leq \|\bh_k^H {\bf f}\|^2, \quad \|{\bf f}\|^2 = 1.
\end{aligned}
\end{equation}
Next, changing the optimization variable from ${\bf f}$ to $\tilde {\bf f} = \frac{1}{\sqrt{\rho}} {\bf f}$, we can simplify the problem and obtain the desired form in \eqref{single_function_AirComp:2}. 
As a result, the solution to \eqref{Problem:lemma:6:1} can be easily derived by scaling the solution to \eqref{single_function_AirComp:2} to meet the unit-norm constraint, which completes the proof.

\bibliography{ref.bib}

\begin{thebibliography}{10}
\providecommand{\url}[1]{#1}
\csname url@samestyle\endcsname
\providecommand{\newblock}{\relax}
\providecommand{\bibinfo}[2]{#2}
\providecommand{\BIBentrySTDinterwordspacing}{\spaceskip=0pt\relax}
\providecommand{\BIBentryALTinterwordstretchfactor}{4}
\providecommand{\BIBentryALTinterwordspacing}{\spaceskip=\fontdimen2\font plus
\BIBentryALTinterwordstretchfactor\fontdimen3\font minus
  \fontdimen4\font\relax}
\providecommand{\BIBforeignlanguage}[2]{{%
\expandafter\ifx\csname l@#1\endcsname\relax
\typeout{** WARNING: IEEEtran.bst: No hyphenation pattern has been}%
\typeout{** loaded for the language `#1'. Using the pattern for}%
\typeout{** the default language instead.}%
\else
\language=\csname l@#1\endcsname
\fi
#2}}
\providecommand{\BIBdecl}{\relax}
\BIBdecl

\bibitem{AgiwalTutorial2016}
M.~Agiwal, A.~Roy, and N.~Saxena, ``Next generation {5G} wireless networks: A
  comprehensive survey,'' \emph{IEEE Commun. Surveys Tuts.}, vol.~18, no.~3,
  pp. 1617--1655, thirdquarter 2016.

\bibitem{zhu2018inference}
G.~Zhu, S.-W. Ko, and K.~Huang, ``Inference from randomized transmissions by
  many backscatter sensors,'' \emph{IEEE Trans. Wireless Commun.}, vol.~PP,
  no.~99, pp. 1--1, Feb. 2018.

\bibitem{KatabiAirComp2016}
\BIBentryALTinterwordspacing
O.~Abari, H.~Rahul, and D.~Katabi, ``Over-the-air function computation in
  sensor networks,'' \emph{CoRR}, vol. abs/1612.02307, 2016. [Online].
  Available: \url{http://arxiv.org/abs/1612.02307}
\BIBentrySTDinterwordspacing

\bibitem{GastparTIT2007}
B.~Nazer and M.~Gastpar, ``Computation over multiple-access channels,''
  \emph{IEEE Trans. Inf. Theory}, vol.~53, no.~10, pp. 3498--3516, 2007.

\bibitem{zhang2006hot}
S.~Zhang, S.~C. Liew, and P.~P. Lam, ``Hot topic: {P}hysical-layer network
  coding,'' in \emph{Proceedings of the 12th annual international conference on
  Mobile computing and networking}.\hskip 1em plus 0.5em minus 0.4em\relax Los
  Angeles, California, USA.: ACM, Sep. 2006, pp. 358--365.

\bibitem{GastparTIT2008}
M.~Gastpar, ``Uncoded transmission is exactly optimal for a simple {G}aussian
  sensor network,'' \emph{IEEE Trans. Inf. Theory}, vol.~54, no.~11, pp.
  5247--5251, Nov 2008.

\bibitem{SoundararajanTIT2012}
R.~Soundararajan and S.~Vishwanath, ``Communicating linear functions of
  correlated {G}aussian sources over a {MAC},'' \emph{IEEE Trans. Inf. Theory},
  vol.~58, no.~3, pp. 1853--1860, March 2012.

\bibitem{GoldsmithTSP2008}
J.~J. Xiao, S.~Cui, Z.~Q. Luo, and A.~J. Goldsmith, ``Linear coherent
  decentralized estimation,'' \emph{IEEE Trans. Signal Process.}, vol.~56,
  no.~2, pp. 757--770, Feb 2008.

\bibitem{C.H.WangTSP2011}
C.~H. Wang, A.~S. Leong, and S.~Dey, ``Distortion outage minimization and
  diversity order analysis for coherent multiaccess,'' \emph{IEEE Trans. Signal
  Process.}, vol.~59, no.~12, pp. 6144--6159, Dec 2011.

\bibitem{GoldenbaumICC2015}
M.~Goldenbaum, S.~Sta{\'n}czak, and H.~Boche, ``On achievable rates for analog
  computing real-valued functions over the wireless channel,'' in \emph{2015
  IEEE International Conference on Communications (ICC)}, June 2015, pp.
  4036--4041.

\bibitem{GoldenbaumWCL2014}
M.~Goldenbaum and S.~Stanczak, ``On the channel estimation effort for analog
  computation over wireless multiple-access channels,'' \emph{IEEE Wireless
  Commun. Lett.}, vol.~3, no.~3, pp. 261--264, June 2014.

\bibitem{GoldenbaumTCOM2013}
------, ``Robust analog function computation via wireless multiple-access
  channels,'' \emph{IEEE Trans. Commun.}, vol.~61, no.~9, pp. 3863--3877,
  September 2013.

\bibitem{GoldenbaumTSP2013}
M.~Goldenbaum, H.~Boche, and S.~Sta{\'n}czak, ``Harnessing interference for
  analog function computation in wireless sensor networks,'' \emph{IEEE Trans.
  Signal Process.}, vol.~61, no.~20, pp. 4893--4906, Oct 2013.

\bibitem{KatabiAirshare2015}
O.~Abari, H.~Rahul, D.~Katabi, and M.~Pant, ``Airshare: Distributed coherent
  transmission made seamless,'' in \emph{Proc. IEEE Conference on Computer
  Communications (INFOCOM)}, Apr. 2015, pp. 1742--1750.

\bibitem{SiggICIT2012}
S.~Sigg, P.~Jakimovski, and M.~Beigl, ``Calculation of functions on the
  {RF}-channel for {IoT},'' in \emph{Proc. IEEE International Conference on the
  Internet of Things}, Oct. 2012, pp. 107--113.

\bibitem{Goldenbaumsensors2014}
A.~Kortke, M.~Goldenbaum, and S.~Sta{\'n}czak, ``Analog computation over the
  wireless channel: A proof of concept,'' in \emph{Proc. IEEE SENSORS}, Nov.
  2014, pp. 1224--1227.

\bibitem{GastparTIT2011}
B.~Nazer and M.~Gastpar, ``Compute-and-forward: Harnessing interference through
  structured codes,'' \emph{IEEE Trans. Inf. Theory}, vol.~57, no.~10, pp.
  6463--6486, Oct. 2011.

\bibitem{T.YangTWC2014}
T.~Yang and I.~B. Collings, ``On the optimal design and performance of linear
  physical-layer network coding for fading two-way relay channels,'' \emph{IEEE
  Trans. Wireless Commun.}, vol.~13, no.~2, pp. 956--967, Feb. 2014.

\bibitem{L.ShiTIT2016}
L.~Shi, S.~C. Liew, and L.~Lu, ``On the subtleties of $q$-pam linear
  physical-layer network coding,'' \emph{IEEE Trans. Inf. Theory}, vol.~62,
  no.~5, pp. 2520--2544, May 2016.

\bibitem{GastparTIT2014}
J.~Zhan, B.~Nazer, U.~Erez, and M.~Gastpar, ``Integer-forcing linear
  receivers,'' \emph{IEEE Trans. Inf. Theory}, vol.~60, no.~12, pp. 7661--7685,
  Dec. 2014.

\bibitem{SakzadTWC2013}
A.~Sakzad, J.~Harshan, and E.~Viterbo, ``Integer-forcing {MIMO} linear
  receivers based on lattice reduction,'' \emph{IEEE Trans. Wireless Commun.},
  vol.~12, no.~10, pp. 4905--4915, Oct. 2013.

\bibitem{GastparProceedings2011}
B.~Nazer and M.~Gastpar, ``Reliable physical layer network coding,''
  \emph{Proceedings of the IEEE}, vol.~99, no.~3, pp. 438--460, Mar. 2011.

\bibitem{gubbi2013internet}
J.~Gubbi, R.~Buyya, S.~Marusic, and M.~Palaniswami, ``Internet of things
  {(IoT)}: A vision, architectural elements, and future directions,''
  \emph{Future generation computer systems}, vol.~29, no.~7, pp. 1645--1660,
  Sep. 2013.

\bibitem{ganti2011mobile}
R.~K. Ganti, F.~Ye, and H.~Lei, ``Mobile crowdsensing: current state and future
  challenges,'' \emph{IEEE Commun. Mag.}, vol.~49, no.~11, pp. 32--39, Nov.
  2011.

\bibitem{foschini1996layered}
G.~J. Foschini, ``Layered space-time architecture for wireless communication in
  a fading environment when using multi-element antennas,'' \emph{Bell Labs
  Technical Journal}, vol.~1, no.~2, pp. 41--59, 1996.

\bibitem{gesbert2007shifting}
D.~Gesbert, M.~Kountouris, R.~W. Heath~Jr, C.-B. Chae, and T.~Salzer,
  ``Shifting the {MIMO} paradigm,'' \emph{IEEE Signal Process. Mag.}, vol.~24,
  no.~5, pp. 36--46, Sep. 2007.

\bibitem{spencer2004introduction}
Q.~H. Spencer, C.~B. Peel, A.~L. Swindlehurst, and M.~Haardt, ``An introduction
  to the multi-user {MIMO} downlink,'' \emph{IEEE Commun. Mag.}, vol.~42,
  no.~10, pp. 60--67, Oct. 2004.

\bibitem{zhu2017hybrid}
G.~Zhu, K.~Huang, V.~K. Lau, B.~Xia, X.~Li, and S.~Zhang, ``Hybrid beamforming
  via the {K}ronecker decomposition for the millimeter-wave massive {MIMO}
  systems,'' \emph{IEEE J. Sel. Areas Commun.}, vol.~35, no.~9, pp. 2097--2114,
  Jun. 2017.

\bibitem{jindal2004duality}
N.~Jindal, S.~Vishwanath, and A.~Goldsmith, ``On the duality of {G}aussian
  multiple-access and broadcast channels,'' \emph{IEEE Trans. Inf. Theory},
  vol.~50, no.~5, pp. 768--783, May 2004.

\bibitem{Weiyu2006duality}
W.~Yu, ``Uplink--downlink duality via minimax duality,'' \emph{IEEE Trans. Inf.
  Theory}, vol.~52, no.~2, pp. 361--374, Feb. 2006.

\bibitem{bjornson2014optimal}
E.~Bjornson, M.~Bengtsson, and B.~Ottersten, ``Optimal multiuser transmit
  beamforming: A difficult problem with a simple solution structure [lecture
  notes],'' \emph{IEEE Signal Process. Mag.}, vol.~31, no.~4, pp. 142--148,
  Jul. 2014.

\bibitem{love2005limited}
D.~J. Love and R.~W. Heath, ``Limited feedback unitary precoding for spatial
  multiplexing systems,'' \emph{IEEE Trans. Inf. Theory}, vol.~51, no.~8, pp.
  2967--2976, Aug. 2005.

\bibitem{choi2006interpolation}
J.~Choi, B.~Mondal, and R.~W. Heath, ``Interpolation based unitary precoding
  for spatial multiplexing {MIMO-OFDM} with limited feedback,'' \emph{IEEE
  Trans. Signal Process.}, vol.~54, no.~12, pp. 4730--4740, Dec. 2006.

\bibitem{peters2011cooperative}
S.~W. Peters and R.~W. Heath, ``Cooperative algorithms for {MIMO} interference
  channels,'' \emph{IEEE Transactions on Vehicular Technology}, vol.~60, no.~1,
  pp. 206--218, Jan. 2011.

\bibitem{medra2015incremental}
A.~Medra and T.~N. Davidson, ``Incremental grassmannian feedback schemes for
  multi-user {MIMO} systems.'' \emph{IEEE Trans. Signal Process.}, vol.~63,
  no.~5, pp. 1130--1143, Mar. 2015.

\bibitem{love2008overview}
D.~J. Love, R.~W. Heath, V.~K. Lau, D.~Gesbert, B.~D. Rao, and M.~Andrews, ``An
  overview of limited feedback in wireless communication systems,'' \emph{IEEE
  J. Sel. Areas Commun.}, vol.~26, no.~8, 2008.

\bibitem{huang2012stability}
K.~Huang and V.~K. Lau, ``Stability and delay of zero-forcing {SDMA} with
  limited feedback,'' \emph{IEEE Trans. Inf. Theory}, vol.~58, no.~10, pp.
  6499--6514, Oct. 2012.

\bibitem{edelman1998geometry}
A.~Edelman, T.~A. Arias, and S.~T. Smith, ``The geometry of algorithms with
  orthogonality constraints,'' \emph{SIAM journal on Matrix Analysis and
  Applications}, vol.~20, no.~2, pp. 303--353, Oct. 1998.

\bibitem{sidiropoulos2006transmit}
N.~D. Sidiropoulos, T.~N. Davidson, and Z.-Q. Luo, ``Transmit beamforming for
  physical-layer multicasting,'' \emph{IEEE Trans. Signal Process.}, vol.~54,
  no.~6, pp. 2239--2251, 2006.

\bibitem{luo2010semidefinite}
Z.-Q. Luo, W.-K. Ma, A.~M.-C. So, Y.~Ye, and S.~Zhang, ``Semidefinite
  relaxation of quadratic optimization problems,'' \emph{IEEE Signal Process.
  Mag.}, vol.~27, no.~3, pp. 20--34, May 2010.

\bibitem{zhang2006eigenvalue}
F.~Zhang and Q.~Zhang, ``Eigenvalue inequalities for matrix product,''
  \emph{IEEE Trans. Auto. Control}, vol.~51, no.~9, pp. 1506--1509, Sep. 2006.

\bibitem{lutkepohlhandbook}
H.~L{\"u}tkepohl, ``Handbook of matrices.'' \emph{John Wiley \& Sons}, 1997.

\end{thebibliography}
\bibliographystyle{IEEEtran}

\end{document}